\documentstyle[preprint,revtex]{aps}
\begin{document}
\font\fortssbx=cmssbx10 scaled \magstep2
\hbox{ 
\fortssbx University of Wisconsin - Madison} 
\hfill\vbox{\hbox{\bf MAD/PH/610}\hbox{February 1992}}\par
\thispagestyle{empty}
\begin{title}
{\bf Intermediate and Heavy Higgs Boson Physics\\ at a 0.5~TeV
{\boldmath $e^+e^-$} Collider}
\end{title}
\author{V.~Barger,$^*$ \ Kingman~Cheung,$^*$ \
B.~A.~Kniehl,$^\dagger$\\ and R.~J.~N.~Phillips$^\ddagger$}
\begin{instit}
$^*$Physics Department, University of Wisconsin, Madison, WI 53706, USA\\
$^\dagger$Deutsches Elektronen-Synchrotron, D-2000 Hamburg 52, Germany\\
$^\ddagger$Rutherford Appleton Laboratory, Chilton, Didcot, Oxon, OX11 0QX,
England
\end{instit}
\begin{abstract}
\nonum
\section{Abstract}
We explore the potential of a future $e^+e^-$ collider in the 0.5~TeV
center-of-mass energy  range to detect  intermediate or heavy Higgs bosons in
the Standard Model. We first briefly assess the production cross sections and
update the decay branching fractions for a Higgs boson of intermediate mass,
with $M_Z<m_H<2M_W$. We then study in detail the possibility of detecting a
heavy Higgs boson, with $m_H>2M_W$, through the production of pairs of weak
bosons. We quantitatively analyze the sensitivity of the process
$e^+e^-\rightarrow \nu\bar\nu W^+W^-(ZZ)$ to the presence of  a heavy
Higgs-boson resonance in the Standard Model. We compare this signal to various
backgrounds and to the smaller signal from $e^+e^-\rightarrow
ZH\rightarrow\mu^+\mu^-W^+W^-(ZZ)$, assuming the weak-boson pairs to be
detected and measured in their dominant hadronic decay modes
$W^+W^-(ZZ)\rightarrow4$\,jets. A related Higgs-boson signal in 6-jet final
states is also estimated.  We show how the main backgrounds from
$e^+e^-W^+W^-\,(ZZ)$, $e\nu WZ$, and $t\bar t$ production can be reduced by
suitable acceptance cuts. Bremsstrahlung and typical beamstrahlung  corrections
are calculated. These corrections reduce Higgs-boson production by scattering
mechanisms but increase production by annihilation mechanisms; they also smear
out some dynamical features such as Jacobian peaks in $p_T(H)$.  With all these
corrections included, we conclude that it should be possible to detect a heavy
Higgs-boson  signal in the $\nu\bar\nu W^+W^-(ZZ)$ channels up to mass
$m_H=350$~GeV.
\end{abstract}

\newpage
\section{INTRODUCTION}
\label{intro}

The structure of the electroweak symmetry-breaking mechanism is a fundamental
issue in particle physics today. A promising way to probe this structure is to
study the production of pairs of $W$ and $Z$ bosons at high-energy
colliders, since the longitudinally polarized states $W_L$ and $Z_L$ derive
their origins and interactions from the symmetry-breaking sector; in
particular, heavy Higgs bosons may appear as resonances. The prospects
for detecting heavy Higgs bosons (with $m_H>2M_Z$) have been analyzed
extensively \cite{snow} for future $pp$ supercolliders. In the present paper we
address the problem of detecting Higgs bosons at
 possible future $e^+e^-$ colliders with
center-of-mass (CM) energies in the 0.5~TeV range, where the experimental
conditions would arguably be  cleaner.

In the Standard Model (SM) there is a single scalar Higgs boson $H$, and the
principal $e^+e^-$ production channels are
\begin{mathletters}
\begin{eqnarray}
e^+e^- \rightarrow & Z^* & \rightarrow ZH \qquad  (ZH\ \rm production)\;,
\label{eq:ee-prod-a} \\
e^+e^- \rightarrow & \nu\bar\nu (W^+)^*(W^-)^* & \rightarrow \nu\bar\nu H
\qquad
(WW\ \rm fusion)\;,   \label{eq:ee-prod-b}
\end{eqnarray}
\end{mathletters}
for which the Feynman diagrams are shown in Fig.~\ref{ee-prod}.  The cross
section of Eq.~(\ref{eq:ee-prod-a}) peaks near CM energy
$\sqrt s=M_Z+\sqrt2\,m_H$
and falls off with increasing energy due to the $s$-channel dependence of
the $Z$-boson propagator.
Contrariwise, the cross section of
Eq.~(\ref{eq:ee-prod-b}), based on $t$-channel vector-boson exchanges, steadily
increases with energy and eventually
surpasses that of  Eq.~(\ref{eq:ee-prod-a}). The $ZZ$-fusion mechanism
analogous to Eq.~(\ref{eq:ee-prod-b}) has weaker couplings and hence a much
smaller cross section.  Figure~\ref{s-depend} shows the various cross sections
versus $\sqrt s$ for $m_H=100$ and $150$~GeV. Results for other $m_H$ values
can be found in Ref.~\cite{barger}. It is interesting to note that
below the nominal $ZH$-production threshold the $ZH$ mechanism (with
a subsequently decaying virtual $Z$ boson) is nevertheless dominant.

At LEP~2, operating at $\sqrt s=200$~GeV, the Higgs discovery limit will be
$m_H\alt 90$~GeV assuming high luminosity \cite{KBG}. An
important task of the Next Linear Collider (NLC) is, therefore, the search for
an intermediate-mass Higgs boson (IMH) in the window
$M_Z<m_H<2M_W$. With a total decay width of $\Gamma_H<100$~MeV, an IMH would be
relatively long-lived \cite{snow}. Thus, the cross sections for the
production of
an IMH via Eqs.~(\ref{eq:ee-prod-a}) or (\ref{eq:ee-prod-b}) and its
subsequent decay into a certain channel may be simply obtained by  multiplying
the curves in Fig.~\ref{s-depend} by the respective branching fraction. In
Ref.~\cite{imh} the branching fractions of an IMH were studied in detail.
We now
update that analysis in the following respects. We include electroweak
radiative corrections at one loop  for all fermionic rates \cite{kni}. In
addition, we include second-order QCD corrections for the quarkonic  widths
\cite{gor} adopting the physical input parameter $m_b=4.25$~GeV from
Refs.~\cite{leut}. We also take into account the QCD corrections to the
two-gluon
mode, which are realized by $H\rightarrow gg(g)$ and $H\rightarrow gq\bar q$
and lead to an enhancement by some 60\% \cite{djo}. Our results are illustrated
in Fig.~\ref{branch}. Figures~\ref{s-depend} and \ref{branch} summarize the
opportunities for an IMH search at the NLC (up to initial-state radiation
corrections that we evaluate below).

In the remainder of this work we shall focus attention on the production of a
heavy Higgs boson, with $m_H>2M_W$. The dominant SM decay modes are then into
four quarks via  $H \rightarrow VV \rightarrow  q\bar q\;q\bar q$
(where $V$ denotes a generic weak boson, $V=W,Z$). In our
subsequent discussion it will be assumed that the $VV$ pairs are measured in
the corresponding four-jet final states $VV\rightarrow j_1j_1j_3j_4$, with
invariant masses $m(j_1j_2)=m(j_3j_4)=M_V$.

At first sight, the simplest and cleanest heavy Higgs signal appears to be in
the dimuon plus four-jet channel
\begin{equation}
e^+e^-  \rightarrow ZH  \rightarrow \mu^+ \mu^- VV\;,
\label{eq:ee->mu j's}
\end{equation}
selecting invariant mass $m(\mu^+\mu^-) = M_Z$;  the Higgs-boson resonance
appea
rs
here as a peak in the $m_{VV}$ distribution, but the
event rate is suppressed by the small branching fraction B$(Z\rightarrow
\mu^+\mu^-)=0.034$ \cite{carter}. It is therefore attractive to investigate
also
 the
invisible $Z$ decay to neutrinos, which has six times larger branching
fraction, B$(Z\rightarrow \nu\bar\nu)=0.20$ \cite{carter};
this leads one to consider the
channel
\begin{equation}
e^+e^- \rightarrow \nu \bar\nu VV \;,   \label{eq:ee->nu j's}
\end{equation}
summed over all neutrino flavors, where there is again a resonance peak in
$m_{VV}$.  The Higgs-boson signal in this channel receives contributions
not only from
$ZH$-production but also from the $WW$-fusion mechanism. This is both an
advantage and a challenge; it further enhances the  event rate and also offers
the opportunity to separate and compare these  contributions and hence to
test the relative
strengths of the $HZZ$ and  $HW^+W^-$ couplings, predicted by custodial SU(2)
symmetry.   The $WW$-fusion contribution to $e^+e^- \rightarrow \nu\bar\nu
W^+W^-(ZZ)$ is not gauge-invariant  by itself however; the complete set of
lowest-order Feynman diagrams for this process is contained
in the generic set shown in Fig.~\ref{e->nuW(Z)}.  The Higgs-boson
signal in channel (\ref{eq:ee->nu j's}) thus has a background from the
non-resonant contributions of Fig.~\ref{e->nuW(Z)}; it also gets
backgrounds from the processes
\begin{mathletters}
\label{eq4}
\begin{eqnarray}
e^+e^- &\rightarrow& \ell^+ \ell^- W^+ W^- (ZZ) \;, \\
e^+e^- &\rightarrow& \ell^\pm \nu W^\mp Z \;,
\end{eqnarray}
\end{mathletters}
when one or more charged leptons escape undetected.

Another channel to consider is
\begin{equation}
e^+e^- \rightarrow  ZH \rightarrow  ZVV \rightarrow  6\,\, {\rm jets}
\label{6jets}
\end{equation}
that benefits from a big branching fraction $B(Z\to jj)=0.70$ \cite{carter}.
The Higgs-boson
signal here appears as a peak in the inclusive $m_{VV}$ spectrum, selecting
final states where the jets reconstruct three weak bosons, and summing over
all possible $VV$ pairings.  This
signal is therefore about 20 times bigger than the $\mu\mu jjjj$ signal,
or 3.5 times bigger than the $ZH$ component of the $\nu\bar\nu jjjj$ signal,
and
hence is roughly comparable to the total $\nu\bar\nu jjjj$ signal. On the other
hand, this channel suffers from a large combinatorial background; even if the
assignment of $jj$ pairs to weak bosons in a given event happens to be unique,
we will have to sum over three possible boson pairings, obtaining
signal/background${}=1/2$ at best. Complications due to jet branchings, jet
overlaps, and demarcation between jets are also more critical here than in the
four-jet final states, where only the total hadronic invariant mass is
ultimately interesting. A proper study of this
channel would require detailed jet simulations, that we do not attempt here;
however, we can infer the potential signal approximately from our
$\mu\mu jjjj$ results.

In addition to intrinsic non-resonant backgrounds and the processes of
Eq.~(\ref{eq4}), significant backgrounds are to be expected from
top-quark pair production,
\begin{equation}
e^+e^- \rightarrow t \bar t \rightarrow ( b W^+)\, (\bar b W^- )\, ,
\label{eq6}
\end{equation}
that can fake $\nu \bar\nu VV$, $\mu^+\mu^- VV$, and $ZVV$ final states
through various decay modes of the $b$ quarks and $W$ bosons.  These
backgrounds cannot be precisely predicted until the top quark is discovered,
but for the energies of present interest they depend rather weakly on the
mass $m_t$ in the range $m_t=135 \pm 35$~GeV indicated by SM studies
\cite{amaldi} and consistent with the experimental bound  \cite{cdf} $m_t
>91$~GeV.  For illustration we shall show the case $m_t=150$~GeV.
We ignore backgrounds
arising from QCD interactions, where continuum light-quark or gluon jets fake
$V
\to
jj$ decays; their cross section is suppressed by additional coupling factors
$\alpha_s^2$ and also by the experimental requirement that  $m(jj)=m_V$.

At high-luminosity $e^+e^-$ colliders, the actual CM energy and CM frame in
which the hard $e^+e^-$ collisions of present interest take place are strongly
affected by electromagnetic radiation in the initial state, including both
bremsstrahlung (which is universal) and so-called beamstrahlung effects
\cite{bai} (which
depend on properties of the beams near their intersection).  Any practical
calculation of signals and backgrounds referring to laboratory conditions
should therefore take account of these corrections.

In the present paper we evaluate the SM Higgs signals and  backgrounds in the
channels (\ref{eq:ee->mu j's})--(\ref{eq6}) above, including
the contributions of all Feynman diagrams of Fig.~\ref{e->nuW(Z)}. We discuss
how cleanly the $ZH$-production and $WW$-fusion contributions  can be
separated in the $\nu\bar\nu VV$ channel and how their event rates compare
(after selection cuts) with the  $ZH$-production signals in the
$\mu^+\mu^-VV$ and six-jets channels.  We take the example of an
$e^+e^-$  collider with
CM energy  $\sqrt s = 0.5$\,TeV,  which is at the lower end of the energy range
0.5--2.0\,TeV  currently being considered and is arguably the easiest to
achieve.  We include effects of initial-state electromagnetic
bremsstrahlung  plus beamstrahlung, taking typical examples of beam parameters
for illustration. After incorporating suitable selection cuts and the
initial-state radiation corrections, we conclude that it should be possible
to detect a heavy Higgs-boson signal in the $\nu\bar\nu W^+W^-(ZZ)$ channels
up to mass $m_H=350$~GeV.

\section{WEAK-BOSON PAIR PRODUCTION IN THE STANDARD MODEL}
\label{sec:weak-boson}

We first consider in turn the leading contributions at tree level to the
production of weak-boson pairs plus two leptons.  An array of generic Feynman
diagrams is shown in Fig.~\ref{e->nuW(Z)}; a detailed explanation of the
diagrams for the various processes is given in Appendix~A. The external fermion
lines can be labelled $e$, $\nu$ or $\mu$ and the external weak bosons are
labeled  $W$ or $Z$. When the external particles are specified,
standard selection rules determine the labeling of almost all
the other lines.  However, a virtual neutral-boson line sometimes represents
both $Z$ and $\gamma$, while the flavor of the final neutrinos is sometimes
$\nu_e$ and sometimes to be summed over $\nu_e, \nu_\mu$, and $\nu_\tau$. The
corresponding helicity amplitudes are given in Appendix A, using the
techniques of Ref.~\cite{stange}.  All the electron
masses are kept finite in the formulas and in the calculations
in order to regulate the mass singularities.

The differential cross sections $d\sigma/dm_{VV}$ for the various reactions
are shown for the case $m_H=200$~GeV in Fig.~\ref{distrib}; the total
integrated cross sections are shown versus $m_H$ in Fig.~\ref{mh-nocut}. No
cuts are imposed here except in the $\mu\mu VV$ channel, where we require
\begin{equation}
\left| m\left(\mu^+\mu^-\right) - M_Z \right| < 15\,\rm GeV \;,
\label{mumu-cut}
\end{equation}
since we are interested only in muon pairs from $Z$ decay. Here and in
subsequent figures we omit all smearing effects arising from experimental
resolution in measurements of invariant masses and transverse momenta.

The cross section for $e^+e^- W^+W^-$ production is
exceptionally large due to photon exchange processes.
The $\nu \bar \nu W^+W^-$ channel is particularly sensitive to Higgs-boson
contributions.
The pronounced edges at $m_H=2M_W$ and $m_H=2M_Z$ in Fig.~\ref{mh-nocut}
indicate the onset of the resonance region for the $s$-channel Higgs
exchange.

\subsection{$e^+e^-\rightarrow\nu\bar\nu\, W^+W^-$}
\label{ssec:ee->nunuWW}

The contributing Feynman diagrams are depicted in (a)--(d), (f),
(h)--(o), (p), (q), (s), (t), and (v)--(y) of Fig.~\ref{e->nuW(Z)} for the
``scattering" channel, where fermion lines 1 and 4 are taken to represent
incoming particles, and in
(a), (c), (f), (g), (h), (j), and (l)--(y) of
Fig.~\ref{e->nuW(Z)} for the ``annihilation" channel, where fermion lines 1
and 2 represent incoming particles.
The corresponding helicity amplitudes are given in Appendices~A.1 and A.2,
respectively. Diagrams
(a)--(g) contribute via $W^+W^-\rightarrow
W^+W^-$ scattering, but of these only (a) and (b)
contain Higgs-boson exchanges and represent the dynamics of the
symmetry-breaking sector that we wish to separate. The other diagrams depend on
scattering or bremsstrahlung via the standard gauge couplings.

We stress that both the ``scattering" and ``annihilation" channels of
$e^+e^- \rightarrow \nu \bar \nu W^+ W^-$ contribute to the Higgs-boson
signal.
In Fig.~\ref{distrib} the curve for this process includes both contributions;
their interference is found to be negligible.
Previous calculations at $\sqrt{s}=1.5$~TeV \cite{japan}
neglected all annihilation-channel contributions; this
approximation is not justified at $\sqrt{s}=0.5$~TeV as Fig.~\ref{PTmiss}(a)
illustrates. We note in passing that our calculations of the scattering
contribution at $\sqrt{s}=1.5$~TeV reproduce the results of
Ref.~\cite{japan}.

Figure~\ref{PTmiss}(a) shows separately the ``scattering" and ``annihilation"
contributions to the $d\sigma/d\overlay{/}{p}_T$ distribution versus missing
transverse momentum $\overlay{/}{p}_T$ for $m_H=200$~GeV.
{}From simple kinematics the distributions terminate at
$\left(\overlay{/}{p}_T\right)_{\rm max}=
\left(s-4M_W^2\right)/\left(2\sqrt s\, \right)=224$~GeV.
The location of the Jacobian peak in the annihilation channel corresponds
approximately to the maximum $\overlay{/}{p}_T$ for which the intermediate
$Z$ and $H$ bosons in $e^+e^- \to ZH \to \nu\bar\nu WW$ are both on mass shell,
$\left(\overlay{/}{p}_T\right)_{\rm peak} =
\left\{\left[s-(M_Z+m_H)^2\right]\left[s-(M_Z-m_H)^2\right]/s\right\}^{1/2}/2
=198$~GeV.
For higher $\overlay{/}{p}_T$ one or both bosons are off mass shell,
which explains the sharp fall beyond the peak.
%
%
We conclude that the scattering and
annihilation contributions can be separated by their $\overlay{/}{p}_T$
spectra (or equivalently by their $p_T(VV)$ spectra, since
$|\overlay{/}{p}_T|=|p_T(VV)|$ in the $\nu \bar \nu VV$ final state).
In Fig.~\ref{PTmiss}(b) we also separate the $\overlay{/}{p}_T$ spectrum of
$e^+e^-\rightarrow \nu \bar \nu H$ into the contributions from $WW$ fusion
and $ZH$ associated production; see Fig.~\ref{ee-prod}.
Now the distributions terminate at
$\left(\overlay{/}{p}_T\right)_{\rm max}=
\left(s-M_H^2\right)/\left(2\sqrt s\, \right)=210$~GeV,
but the position of the Jacobian peak in the $ZH$ curve is unchanged.
For $m_H=200$~GeV, a
$\overlay{/}{p}_T>175$~GeV cut can single out the contribution of
the $ZH$ diagram, assuming the CM energy to be 0.5 TeV.
We shall see later that initial-state radiation corrections
smear the CM energy and with it the $\overlay{/}{p}_T$
dependence; nevertheless, it remains true that $ZH$ production and
$VV$ fusion contribute in distinctively different ways and can in principle be
separated by a detailed study of the $\overlay p/_T$ dependence of the signal.

\subsection{$e^+e^-\rightarrow \nu\bar\nu ZZ$}
\label{ssec:ee->nunuZZ}
The contributing Feynman diagrams are depicted in (a), (c), and (f)--(u) of
 Fig.~\ref{e->nuW(Z)} for the scattering channel and in (a), (b), (h)--(k), and
(p)--(u) for the annihilation channel with appropriate particle labeling. The
helicity amplitudes are given in Appendices~A.3 and A.4.
This case too has contributions from ``scattering" and ``annihilation"
channels, and both contain the Higgs-boson signal. However, the cross section
 is much smaller than that for $e^+e^- \rightarrow \nu \bar \nu W^+ W^-$.
The major background in this channel arises in the annihilation channel from
$e^+e^- \rightarrow ZH$ with $H \rightarrow ZZ \rightarrow Z \nu \bar \nu$
decay $\bigl($Fig.~\ref{e->nuW(Z)}(b)$\bigr)$; it comes from the Higgs boson
its
elf,
via a different decay mode from the one we are studying, and gives the
broad enhancement at large $m_{VV}$ in Fig.~\ref{distrib}(a).

\subsection{$e^+e^-\rightarrow e^+e^-W^+W^-$}
\label{ssec:ee->eeWW}
The contributing Feynman diagrams are depicted in (a), (c), (f), (g), (i),
(k), and (l)--(y) of Fig.~\ref{e->nuW(Z)} and explained in Appendix A.5.
Although this channel has
contributions from Higgs-exchange diagrams, it is essentially
a background to the Higgs signal, because
it is overwhelmingly due to $\gamma$-exchange contributions.  In
Figs.~\ref{distrib}(a) and \ref{mh-nocut}(a) we can see
that this background is potentially much bigger than the
signal in the $\nu \bar \nu W^+W^-$ channel.    But with a suitable $p_T(VV)$
cut and central-electron vetoing, we shall show that this
background can be reduced to an unnoticeable level.

\subsection{$e^+e^-\rightarrow e^+e^-ZZ$}
\label{ssec:ee->eeZZ}
The contributing Feynman diagrams are (a), (b), (h)--(k), and
(p)--(u) of Fig.~\ref{e->nuW(Z)} with the
appropriate substitutions (see Appendix A.6).  This channel has
characteristics similar
to II.C, but the cross section is exceedingly small because of the small $Zee$
coupling.

\subsection{$e^+e^-\rightarrow e^\pm \nu W^\mp Z$}
\label{ssec:ee->enuWZ}
The contributing Feynman diagrams are (b)--(r), (t), (u), and (w)--(y) of
Fig.~\ref{e->nuW(Z)}; the corresponding amplitudes are listed in Appendix A.7.
 This process gives another background to the signal; the cross section depends
on $m_H$ because of the Higgs-boson exchange diagram \ref{e->nuW(Z)}(b).
With a $p_T(VV)$ cut and central-electron
vetoing it can be greatly reduced, as explained
below.  Furthermore, if it became possible to distinguish accurately between
$W$ and $Z$ from the invariant masses of their decay dijets, then this
background could be removed.

\subsection{$e^+e^-\rightarrow \mu^+\mu^- W^+W^- (ZZ)$}
 The amplitudes in these cases precisely equal the lepton
annihilation amplitudes for the  $e^+e^- \rightarrow e^+e^-W^+W^-(ZZ)$ cases,
described in II.C and II.D above.  In addition to the
$H \rightarrow W^+W^-(ZZ)$  diagrams that give the Higgs-mass peaks in the
$m_{VV}$  distributions, there are weak-boson scattering and
bremsstrahlung diagrams that contribute backgrounds.

In the  $\mu^+\mu^-ZZ$  channel there is also a contribution from
$e^+e^- \rightarrow ZH$  production with  $H \rightarrow ZZ \rightarrow
\mu^+\mu^-Z$  decays $\bigl($Fig.\ref{e->nuW(Z)}(b)$\bigr)$;
this appears as a background to the Higgs signal in the  $m_{ZZ}$
distribution, if both these $Z$ bosons are assumed to decay
hadronically.  Since we have here a $ZZZ$ final state, the event rate in the
Higgs peak (and background) could be tripled if we included all three $ZZ$
pairings in the $m_{ZZ}$ distribution.  However, this procedure would make
sense only if the hadronically decaying $Z$ bosons could be clearly
distinguished from $W$ bosons (it would be nonsensical in $\mu^+\mu^-W^+W^-$
final states) and we do not pursue it here.

\subsection{$e^+e^- \rightarrow t \bar t \rightarrow b \bar b W^+ W^-
\rightarrow b \bar b f_1 \bar f_2 f_3 \bar f_4$}

We calculate this background process at tree level, including full spin
correlations down to the final $W \rightarrow f_i \bar f_j$ decays
into fermions, following the techniques of Ref.~\cite{stange}.
Very heavy quarks decay before they have time to hadronize or
depolarize \cite{bbhp}, so we treat the  $t\rightarrow bW$  decays at
the quark level with 100\% branching fraction.
When semileptonic $b$ decays are needed, we first hadronize the $b$ quark to a
$B$ hadron in the laboratory frame via the Peterson \cite{peterson} model
with parameter $\epsilon=0.05$, consistent with LEP studies \cite{L3}, and
then use the free-quark $b \rightarrow c \ell \nu$ decay matrix elements
following the spectator model of heavy-quark decays.
The basic $e^+e^- \rightarrow  t\bar t$ cross section is relatively large,
of order 700--500 fb for $m_t = 100$--200~GeV at $\sqrt{s}=500$~GeV.

This final state can be mistaken for $\nu \bar \nu VV$ if (a) one $W$ boson
decays leptonically $W \rightarrow \ell \nu$, (b) the charged lepton escapes
detection while the neutrino gives large $\overlay{/}{p}_T$,
and (c) the final $b \bar b$ pair fakes a $V \rightarrow jj$ decay.
This background can be reduced by requiring
large $\overlay{/}{p}_T$ and by vetoing central leptons, the same
criteria that suppress the other backgrounds.  We note that leptonic $\tau$
decays give final $e$ or $\mu$ while hadronic $\tau$ decays give identifiable
narrow jets; although the technical details for vetoing central taus must
differ from those for vetoing central electrons and muons, we shall here treat
t
hem
all the same as a first approximation.
In Figs.~\ref{distrib}(a) and \ref{mh-nocut}(a) we show the potential $t\bar
t$ background in the $\nu\bar\nu VV$ channel, before making selective cuts, for
the case $m_t = 150$~GeV. We first calculate the cross section
for  $\ell\nu jjjj$  production, summing the contributions from
$\ell=e,\mu,\tau$, and requiring that the fake $V$ generated by
$b$ and $\bar b$ jets has invariant mass within 15 GeV of the $W$ or $Z$ mass:
\begin{equation}
M_W - 15 \,{\rm GeV}  <   m(b \bar b\,)  <   M_Z + 15 \,{\rm GeV}.
\label{eq8}
\end{equation}
This cross section is divided by the square of the branching
fraction $B(V\rightarrow jj)=0.70$  to convert it to an effective $\nu\bar\nu
VV$  cross section. Then $m_{VV}$ denotes the invariant mass of the
hadronically decaying $W$ plus the fake $V$.

These  $t\bar t$ final states faking $\nu\bar\nu VV \rightarrow \nu\bar\nu
jjjj$
all have the property that three of the jets have invariant mass $m(jjj)=m_t$.
We could therefore in principle suppress this background completely by vetoing
f
inal
states where any three jets satisfy  $|m(jjj) - m_t| < 15$~GeV,
say, once the top quark has been discovered and $m_t$ is known.
The price paid would be the loss of some of the $H \rightarrow VV \rightarrow
 jjjj$  signal, but this price varies greatly according
to the values of $m_t$  and  $m_H$.  If $m_t = 150$~GeV, for example,
this invariant mass veto would destroy about 90\% (70\%) of the Higgs-boson
signal for $m_H=200$~GeV (250~GeV), which is probably too much to pay.
But if $m_t = 100$~GeV , the corresponding Higgs signals would
be reduced by only 45\% (20\%) instead, while for $m_H=300$ or 350 GeV
the loss of signal would be less than 10\%.  All this assumes of
course that the eventual detectors would allow sufficiently
accurate multijet invariant mass reconstructions.  In our present
analysis we do not apply this three-jet mass veto, since it would
introduce a confusing multiplicity of cases to discuss; however,
it remains a potentially helpful cut for future application.

Top-quark pair production can also be mistaken for $\mu\mu VV$ production if
two of the heavy flavors decay into muons.  Two primary top semileptonic
decays give the wrong jet multiplicity, but one primary $t$ decay plus one
secondary $b$ decay to muons can fake $\mu\mu VV$ if the hadronic debris from
the two $b$ quarks (after one semileptonic decay) combines to fake
$V \rightarrow jj$.  With smaller probability, two secondary $b$ decays to
muons can also fake $\mu\mu VV$, if the hadronic debris from the $b$ quarks
does not obscure the two remaining genuine $W$ bosons; we shall ignore this
contribution and other small effects arising from semileptonic charm decays.
These backgrounds can
be suppressed by requiring the dimuon invariant mass to be close to $M_Z$ and
by vetoing events with large $\overlay{/}{p}_T$.  They can be suppressed
further by adding some isolation requirement on the muons, since $Z
\rightarrow \mu\mu$ typically gives isolated muons whereas $b$ and $c$ decays
give muons in or near jets, but we do not purse this further here.

In Figs.~\ref{distrib}(b) and \ref{mh-nocut}(b) we show the potential
$t\bar t$ background in the $\mu\mu VV$  channel, before making selective
cuts, for the case $m_t = 150$~GeV. We calculate the cross section
for  $\mu\mu jjjj$  production, requiring that the dimuon invariant
mass satisfies the constraint Eq.~(\ref{mumu-cut}) and that the fake $V$ from
$b+\bar b$ hadronic debris satisfies the constraint Eq.~(\ref{eq8}).
This cross section is divided by the square of the branching
fraction $B(V\rightarrow jj)=0.70$ , to reduce it to an effective  $\mu\mu VV $
cross section.  Then $m_{VV}$ denotes the invariant mass of one
true $W$ plus one fake $V$, as before.

Finally, $t\bar t$ production can also be mistaken for $ZVV\rightarrow$ 6
jets production if
all the $W$ bosons and $B$ hadrons decay hadronically and the two $b$ jets
fake a $V \rightarrow jj$ configuration.
This background is suppressed a little by the decay
branching fractions and rather more by the constraint
$m(b\bar b\,) \approx M_V$.  It could be suppressed completely
by vetoing  $m(jjj) \approx m_t$  configurations, as
discussed above, but we do not apply this veto in our present work.

\section{RESULTS AFTER KINEMATIC CUTS}

At $\sqrt{s}=0.5$~TeV there are seldom any events for $W$ bosons having
absolute
rapidity $|y|>$ 1.5, and there is not much difference between rapidity cuts of
1.0 and 1.5.
Furthermore,
experimental acceptance will force some such cut upon us anyway, since jet
measurements will be impossible close to the beam axis. In our calculations we
shall therefore make the cut
\begin{equation}
|y(V)|<1 \label{rap-cut}
\end{equation}
to approximate the experimental acceptance, with no serious loss in signal.

In the $\nu\bar\nu VV$ channel, the backgrounds from $e^+e^-
\rightarrow e^+e^- W^+W^-$ (mainly due to $\gamma$-exchange processes),
from $e^+e^- \rightarrow e \nu WZ$, and from $e^+e^- \rightarrow t \bar t$
are potentially dangerous;
see Figs.~\ref{distrib} and \ref{mh-nocut}.  Our strategy to reduce the first
two backgrounds is similar
to that in Ref.~\cite{japan}.  We first suppress the contribution from the
 $\gamma$-exchange poles by a cut
\begin{equation}
p_T(VV)>45 \; \mbox{GeV}\;,
\end{equation}
which effectively removes the
double-$\gamma$ pole contribution to $e^+e^- \rightarrow e^+e^-W^+W^-$.
 In addition, we veto events with a visible electron.  We
assume that $e^\pm$ will be identifiable if they have high energy and are
emitted in the central region (neglecting the possibility of losing electrons
in jets); we therefore veto all events that contain
$e^\pm$ with
\begin{equation}
E_{e^\pm} > 50 \;\mbox{GeV \hspace{0.2in}and\hspace{0.2in}}
|\cos(\theta_{e^\pm})| < \cos(0.15) \;. \label{veto}
\end{equation}
This proves to be very effective in reducing the $e^+e^- W^+W^-$ background,
but less in suppressing the $e \nu WZ$ background.
The $t\bar t$ background is little suppressed by the
$p_T(VV)$ cut but is considerably reduced by vetoing
all central leptons; for this we extend the criteria
of Eq.~(\ref{veto}) to apply to all charged leptons $e,\,\mu$, and $\tau$.

The results for $\nu\bar\nu VV$ signal and background channels are summarized
in Fig.~\ref{dist-cut} for the
differential cross section versus the invariant mass $m_{VV}$ of the weak-boson
pair, and in Fig.~\ref{mh-cut} for the total cross section versus $m_H$;
see also Table~I. With the above cuts the total cross
sections for $e^+e^- \rightarrow e^+e^- W^+W^-$ and $e^+e^-ZZ$ are now
exceedingly small, being of order 0.1 and 0.01~fb
respectively. We see that there is an excellent Higgs-boson signal with little
background, especially for the mass range 160~GeV $<m_H<225$~GeV, where the $H
\rightarrow WW,\,ZZ$ resonance signal exceeds 20~fb with a total
non-resonant background of about 40~fb.  For an assumed annual integrated
luminosity of $50~{\rm fb}^{-1}$, two-jet branching fractions $B(W\rightarrow
jj) \approx B(Z\rightarrow jj) =0.7$, and perhaps 50\% instrumental efficiency,
a 20~fb signal translates into 250 signal events per year in the $H\rightarrow
VV\rightarrow 4$\,jets channels. It should be possible to detect much smaller
signals than this.

Further details of the $e^+e^-\rightarrow \nu\bar\nu VV$ signals and
backgrounds
are given in Table~\ref{various-ex}, where total cross sections are listed for
a range of Higgs mass values (excluding bremsstrahlung and beamstrahlung
corrections,
as do all our calculations so far). The intrinsic background in the $\nu
\bar \nu WW$ channel, i.e.\ the cross section without the resonance peak in
$m_{WW}$, is labeled by $\nu\bar\nu WW(0)$ and is estimated by calculating
with some small Higgs-boson mass, $m_H=50$~GeV say; this approximation
agrees closely with the heavy-Higgs
calculations outside the resonance peak and interpolates smoothly under the
peak. Subtracting this background from the total $\nu\bar\nu WW$ cross
section gives an estimate of the Higgs signal in this channel.
The sum of the intrinsic and
other (misidentification) backgrounds gives the total integrated background.
In the $\nu \bar \nu ZZ$ channel, the principal intrinsic background comes
from $ZH$-production with $H \rightarrow Z \nu \bar \nu$ decay, which cannot
be obtained from any light-$H$ calculation; a rough estimate of this
background is provided by taking 2/3 of the integrated $\nu \bar \nu ZZ$
production, with the remaining 1/3 being due to the signal.
Note however that for discussions of the significance of each signal, we should
compare with the part of the background {\em lying under the resonance peak},
that is an order of magnitude smaller than the integrated background in each
channel given by our Tables.

The Higgs-boson signals in the $ZH\rightarrow\mu\mu VV$ channels do not suffer
from
potentially large backgrounds and therefore do not need strong cuts for their
extraction. However, the $t\bar t$ background has typically large
$\overlay{/}{p}_T=p_T(VV)$ due to the leptonic $W$-boson decay, unlike the
signal or the other backgrounds; we therefore choose to suppress $t\bar t$
contributions by requiring
\begin{equation}
\overlay{/}{p}_T < 40 \,{\rm GeV}\;.
\label{pt40}
\end{equation}
In addition to the $\mu^+\mu^-$ mass constraint of Eq.~(\ref{mumu-cut}) and
the fake-$V$ mass constraint of Eq.~(\ref{eq8}), we also impose the rapidity
cut of Eq.~(\ref{rap-cut}),
approximating the likely experimental acceptance, both on the vector bosons and
on the $\mu^+\mu^-$ system.  The total cross section and $m_{VV}$ dependence
differ little from the uncut distributions shown in Figs.~\ref{distrib}(b)
and \ref{mh-nocut}(b) (apart from $t\bar t$ contributions), so we do
not plot them  again. Details of the integrated  $\mu\mu VV$   signals and
backgrounds are given in Table~\ref{mumu-ex}.  In the $\mu\mu WW$ channel the
background $\mu\mu WW(0)$ is estimated by calculating with a light Higgs
boson, as in the $\nu \bar \nu WW$ channel (Table~I).  In
the $ZZZ \rightarrow \mu\mu ZZ$ channel, where the final $ZZ$ are assumed to
decay into four jets, the principal background comes from $H \rightarrow ZZ
\rightarrow \mu\mu jj$ decays; this accounts for about 2/3 of the integrated
cross section while the remaining 1/3 comes mostly from the $H \rightarrow ZZ
\rightarrow jjjj$ signal.  We note that the net signal cross sections in the
$\mu\mu VV$
channels (Table~\ref{mumu-ex}) are about 10--20 times smaller than those in the
$\nu \bar \nu VV$ channels (Table~\ref{various-ex}).

The net Higgs signal in the $ZH\to 6$ jets  channel is related to that in
the $ZH\to \mu\mu jjjj$ channel by the ratio of branching fractions $B(Z\to
jj)/B(Z\to\mu^+\mu^-) \approx 20$, as remarked in Section~I, if we continue to
apply essentially the same acceptance cuts both to the muons and to the jets.
We can therefore use the latter signal to estimate the former; folding in the
$V \to jj$ branching fractions, we see from Table~II that the Higgs-boson
signal in 6-jet final states falls from about 15~fb at $m_H=175$~GeV to
about 3~fb at $m_H=350$~GeV.  The $e^+e^-
\rightarrow VVV \rightarrow 6$ jets background is harder to estimate however;
it depends on detailed questions of jet resolution, etc. However, the minimal
combinatorial background from the $ZH \rightarrow ZVV$ process is at least
twice the signal.  Also, we calculate that the effective $t\bar t \rightarrow
VVV\to$ 6 jets background (with $b\bar b$ faking a $V$) is of order 75~fb for
$m_t=150$~GeV, if we simply impose the mass and rapidity cuts of
Eqs.~(\ref{eq8}) and (\ref{rap-cut}).  Compared to the $\nu\bar\nu VV$ case,
the signals in this channel have similar strength but the backgrounds are
much bigger and more problematical.

Finally we recall that the effects of experimental resolution, arising from
calorimeter fluctuations, semileptonic decays in jets, etc., remain to be added
to our present calculations. These effects will somewhat smear the sharp peaks
in invariant masses and transverse momenta that we show. There are also
important corrections from initial-state radiation, that we now discuss.

\section{EFFECTS OF BREMSSTRAHLUNG AND BEAMSTRAHLUNG}
\label{sec:brems-beam}

It is well known that cross sections measured in high-energy $e^+e^-$
collisions are greatly affected by QED radiative corrections.
(However, we can
as a first approximation neglect weak corrections, which are known to vary
between $-7\%$ and $+6\%$ for the $e^+e^- \rightarrow ZH$ cross section at
$\sqrt{s}=0.5$~TeV \cite{kni2}.) In general, the
main effect is due to bremsstrahlung from the initial state, which lowers the
effective CM energy available in the main process and leads to a
typical  smearing of the distributions in the subprocess CM energy
$\sqrt{\hat s}$. The size of the effect can be
easily estimated by considering the large  leading logarithms. In
${\cal O}(\alpha^n)$ they are of the form $(\alpha/\pi)^n\ln^n(s/m_e^2)$
$(n=1,2,\ldots)$, which follows from Sudakov's theorem \cite{sud}. In an
inclusive experiment there are no similar logarithms due to final-state
particles. This is guaranteed by the Kinoshita-Lee-Nauenberg theorem
\cite{kin}, which states that mass singularities associated with outgoing
particles cancel when all final states with the same invariant mass are summed
up. For $\sqrt s=0.5{\rm\, TeV}$ one has $(\alpha/\pi)\ln(s/m_e^2)\approx7\%$.
It is, therefore, clear that any predictions for processes at the NLC that
ignore initial-state bremsstrahlung will vary from crude to
completely inadequate.
By analogy to the situation at the $Z$ peak it is clear that
in the presence of a not-too-broad Higgs resonance even
a rigorous treatment to ${\cal O}(\alpha)$ would still fail to lead to
reliable predictions.
An ${\cal O}(\alpha^2)$ calculation in connection with soft-photon
exponentiation \cite{yen} would be in order.
Unfortunately, full ${\cal O}(\alpha^2)$ results are only available for the
class of processes where the electron-positron pair annihilates into a
neutral gauge boson \cite{ber}.
However, the pattern in which the leading logarithms arrange themselves
is universal for all reactions with an $e^+e^-$ initial state.
Leading-order initial-state bremsstrahlung is most conveniently included by
convoluting the reduced cross section with exponentiated splitting functions
for the incident electron and positron beams \cite{kur}.
These splitting functions characterize the probability of finding an
electron (positron) with a given longitudinal-momentum fraction
inside the original electron (positron) after multiple-photon emission
and can be obtained by solving QCD-like evolution equations.
In our analysis we adopt Eq.~(20) of the first paper of Ref.~\cite{kur}
neglecting the numerically insignificant contribution of those $\beta^2$
terms which do not participate in the exponentiation.
For $e^+e^-$ reactions in the
continuum it is evident that initial-state bremsstrahlung should tend to reduce
(enhance) a cross section which is increasing (decreasing) with increasing
energy.

A completely novel feature which will be faced at the NLC is the
so-called beamstrahlung phenomenon \cite{bai},
which is an unavoidable consequence of the quest for luminosities
which exceed current achievements by three orders of magnitude.
It occurs when particles in one bunch undergo bremsstrahlung
upon entering the electromagnetic field of the other bunch.
These particles thus interact coherently with a sizeable part
of the opposite bunch.
The intensity of the emitted beamstrahlung therefore increases with
the strength of the fields generated by the bunches, which in turn
grows with the particle density of the bunches and hence with the
luminosity per bunch crossing.
Beamstrahlung effects delicately depend on the details of the machine
operation
and a realistic estimate of their characteristics can only be obtained
through Monte Carlo simulation.
In our analysis we generate beamstrahlung (and bremsstrahlung) events using
the program package BEAMSPEC by Barklow \cite{tim}.

The optimization of a linear-collider design proceeds in a multi-dimensional
parameter space with a network of constraints \cite{palmer}.
The currently existing concepts for a 500~GeV NLC fall into three
broad categories \cite{igo}:
(1) SLAC and KEK propose a traveling-wave copper structure at
room-temperature, operating at 11.4~GHz (X-band),
and a gradient of 50--100~MV/m.
There is an option with high luminosity but strong beamstrahlung (Palmer~G)
and an alternative with moderate beamstrahlung at the cost of a factor 2--4
in luminosity (Palmer~F) \cite{palmer}.
(2) DESY/Darmstadt proposals extend the present SLC technology to higher
energies, using a warm travelling-wave copper structure, operating at
2.8~GHz, and a gradient of 17~MV/m.
The original wide-band version \cite{voss} suffers from a large intrinsic
linac energy spread, which impairs the resolving power for very narrow
structures.
This drawback has been largely removed in the narrow-band version
\cite{weiland}.
(3) The TESLA design \cite{wiik} proposes to use a superconducting
standing-wave

radio-frequency structure at 1.3~GHz with a gradient of 25~MV/m.
Thanks to very long bunches and large spot sizes, beamstrahlung
is reduced to a level comparable to initial-state bremsstrahlung.

It has become customary \cite{noble} to characterize the beamstrahlung
spectrum by a dimensionless {\it beamstrahlung parameter} $\Upsilon$,
which is defined by $\Upsilon=\gamma B/B_c$,
where $\gamma=E_b/m_ec^2$ is the ratio of the initial beam energy to the
electron mass,
$B$ is the average magnetic field inside a bunch,
and $B_c=m_e^2c^3/e\hbar=4.4\times10^{13}$~Gauss is the Schwinger critical
field.
For $\Upsilon<1$, the beamstrahlung energy loss is a monotonically
increasing function of $\Upsilon$ \cite{palmer}.
For the above designs the numbers are 0.385 (Palmer~G), 0.108 (Palmer~F),
0.065 (DDwb), 0.013 (DDnb), and 0.008 (TESLA) \cite{igo}.
To start with, we consider the Palmer~G design to mark the most
disadvantageous scenario.
Figures \ref{new10}--\ref{103} and Tables III, IV display the characteristic
fea
tures
of bremsstrahlung and beamstrahlung for a collider of this type.
For comparison, we then repeat the central parts of our analysis assuming
the more favorable DESY/Darmstadt narrow-band design;
see Figs.~\ref{fourteen}, \ref{fifteen} and Tables V, VI.
The corresponding results for TESLA are very similar to those
for DESY/Darmstadt.

We first illustrate the implications for intermediate- and high-mass
Higgs-boson
production rates. Figure~\ref{new10} shows the $e^+e^- \rightarrow \nu\bar\nu
H$ cross sections via the
$ZH$-production and $WW$-fusion subprocesses versus $m_H$ at $\sqrt
s=0.5$~TeV, before and after corrections for bremsstrahlung and beamstrahlung.
We see that these corrections enhance the $ZH$-production mechanism for $m_H
\alt 200$~GeV; here the increase in cross section from smearing
to lower CM energy wins over threshold effects, but for higher $m_H$ the
threshold suppression takes over.  Production via $WW$-fusion is reduced for
all $m_H$ values, since the cross section falls monotonically as
$\sqrt{\hat s}$ is decreased.

Figure~\ref{101} illustrates some  characteristic features of
beamstrahlung and bremsstrahlung on the $\nu\bar \nu H$ signals
from $WW$-fusion and $ZH$ associated production,
assuming $m_H=200$~GeV.  Figure \ref{101}(a)
shows the $\sqrt{\hat s}$ dependence after these
corrections are taken into account;
without corrections the entire process would occur at $\sqrt{\hat
s}=0.5$~TeV. As expected, the $WW$-fusion process, which has a cross section
rising with $\hat s$, contributes mostly at the top of the $\hat s$ range; the
$ZH$-production process, with a falling cross section,
contributes across a wide range of $\hat s$.
Figure~\ref{101}(b) shows the $\overlay{/}{p}_T$ distributions before and
after corrections; we see that contributions from the upper $\overlay{/}{p}_T$
range (that require close-to-maximum values of $\sqrt{\hat s}\,$)
are strongly suppressed, and the Jacobian peak near maximum
$\overlay{/}{p}_T$ is smeared out.

In Fig.~\ref{102} we investigate the the impact of bremsstrahlung and
beamstrahl
ung
on $e^+e^-\to\nu\bar\nu W^+W^-$ differential cross sections,
distinguishing between the scattering and annihilation channels;
this contains the Higgs-boson signals previously shown in Fig.~\ref{101},
but now the intrinsic backgrounds are also present.
Figures \ref{102}(a) and (d) display the $\sqrt{\hat s}$ dependences.
Figures \ref{102}(b) and (e) show the distributions versus invariant mass
$m_{WW
}$;
obviously, the smearing in $\hat s$ has no effect on locations and widths
of the Higgs resonance peaks, but the height of each peak and the
distribution of its intrinsic background can be changed.
Figures \ref{102}(c) and (f) show the smearing of the $\overlay{/}{p}_T$
distrib
utions.

Tables \ref{various-in} and \ref{mumu-in} give the final cross sections for
the signals and backgrounds in the $\nu \bar \nu VV$ and $\mu\mu VV$
channels, after adding bremsstrahlung and Palmer G-type beamstrahlung
corrections to the previous results shown in
Tables \ref{various-ex} and \ref{mumu-ex}, respectively; we recall that this
is the worst beamstrahlung scenario. In the
$\nu \bar\nu VV$ channels, the Higgs signal is appreciably reduced while the
background from $e \nu WZ$ is somewhat less reduced (or even
increased).  The $t\bar t$ background is doubled, partly because $t\bar t$
production is increased but mostly because the cuts become less effective at
lower $\sqrt{\hat s}$.  Nevertheless, a healthy signal remains right through
the
mass range illustrated here. For $m_H=300\ (350)$~GeV the total signal would
 amount to 2.7 (0.8)~fb compared to a total background of 40 (39)~fb,
of which the component under the mass peak would be only about 4 (1.3)~fb;
see Fig.~\ref{103}.
If we assume as before an annual integrated luminosity of $50\,{\rm fb}^{-1}$
and 50\% instrumental efficiency and take dijet branching fractions $B(V
\rightarrow jj)=0.7$, then for $m_H=300$~GeV there would be about 30
events/year in the combined $\nu\bar\nu VV \rightarrow \nu\bar\nu jjjj$
channels, with a background of about 50 events/year in the mass bin below the
Higgs resonance peak.  For $m_H=350$~GeV there would be about 10 signal
events/year, with about 15 background events/year below the peak.  It
appears that a Higgs signal would be detectable in this channel for a
mass up to $m_H = 300$~GeV quite readily, and up to $m_H=350$~GeV eventually.
The net Higgs-boson signal in the $\mu\mu VV $ channels is smaller by a factor
of 15--40, depending on $m_H$.

Figure~\ref{103} shows the total Higgs signals and backgrounds
in the $\nu\bar\nu VV$ and $\mu\mu VV$ channels, versus invariant mass
$m_{VV}$. All the cuts (\ref{eq8})--(\ref{veto}) for $\nu\bar\nu VV$
and (\ref{mumu-cut}), (\ref{eq8}), (\ref{rap-cut}), (\ref{pt40}) for $\mu\mu
VV$ have been applied here, and all bremsstrahlung and
Palmer~G beamstrahlung corrections have been made. In Fig.~\ref{103}(a) all
the signals and backgrounds for $\nu\bar\nu VV$ channels are shown for
$m_H=200$~GeV. In
Fig.~\ref{103}(b) we show the sum of $\nu\bar\nu WW$ and $\nu\bar\nu ZZ$
differe
ntial
cross sections, since these channels would be practically
indistinguishable in their hadronic decay modes $VV \rightarrow jjjj$, for
$m_H=200$, 250, 300, and 350 GeV.
At the same time the major
backgrounds from $e^+e^- \rightarrow e \nu WZ$ and from $e^+e^- \rightarrow
t\bar t$, for the same choices of $m_H$,
are added to the corresponding Higgs-boson signal.  Similarly,
Fig.~\ref{103}(c) shows the combined differential cross sections of the
$\mu\mu WW$ and $\mu\mu ZZ$ signals and the $t\bar t$ background
for the same choices of $m_H$. This figure neatly summarizes our
principal heavy-Higgs results in the Palmer~G case.

We now examine the performances of the DESY/Darmstadt
and TESLA designs with regard to beamstrahlung suppression.
In Fig.~\ref{fourteen} we compare for $m_H=200$~GeV the $\sqrt{\hat s}$
dependen
ces of
the differential cross sections $d\sigma/d\sqrt{\hat s}$ for $\nu\bar\nu H$
production which arise from Palmer~G and DESY/Darmstadt (narrow band).
We consider separately the contributions from $WW$-fusion in
Fig.~\ref{fourteen}
(a)
and from $ZH$ associated production in Fig.~\ref{fourteen}(b).
The curves for Palmer~G are the same as in Fig.~\ref{101}(a).
We see that the distributions for DESY/Darmstadt are much more
massed close to the nominal value $\sqrt s=500$~GeV, i.e.\ the average
energy loss of the incident beams by beamstrahlung is drastically reduced.
In particular, the plateau in the curve of $ZH$-production is absent
for this design.

In Tables~V and VI we repeat the analysis of Tables~III and IV
adopting the DESY/Darmstadt design parameters (the TESLA design gives almost
identical results).  The results lie throughout much closer to the
uncorrected case of Tables~I and II than to the case of Palmer~G.
This is again nicely demonstrated in Table~VII, which
shows how much the total cross sections of $\nu\bar\nu VV$ and
$\mu^+\mu^- VV$ production, with $VV=WW$, $ZZ$, are changed under the
influence of beamstrahlung and bremsstrahlung for the various designs.
We conclude that the smearing effects of beamstrahlung do not necessarily
represent a serious danger for Higgs hunting at a 500~GeV NLC.
It is an issue of machine architecture and operation to reduce
unwanted beamstrahlung to the level of (unavoidable) bremsstrahlung.

Finally, in Fig.~\ref{fifteen} we repeat the analysis of Fig.~\ref{103} for the
DESY/Darmstadt design.
The $m_{VV}$ distributions shown in Fig.~\ref{fifteen}(a) are much closer to
the

uncorrected case of Fig.~\ref{dist-cut} than to the Palmer G case of
Fig.~\ref{1
03}(a).
In particular, the bumps at the upper end of the $m_{VV}$ range
in the distributions of the $eeVV$ $(VV=WW,ZZ)$ channels, which can be traced
to those annihilation diagrams where the $V$ bosons are emitted from the
initial state $\bigl($see Figs.~\ref{e->nuW(Z)}(x) and (y)$\bigr)$, are much
les
s washed
out here than in the case of Palmer~G.
Furthermore, the signal peaks are more prominent than for Palmer~G;
however, as we have noted already in the context of Figs.~\ref{102}(b) and (d),
their locations and widths are insensitive to beamstrahlung.
These observations are substantiated by Figs.~\ref{fifteen}(b) and (c).
The signals are somewhat bigger and the backgrounds are somewhat smaller than
for the Palmer~G design; we conclude once more that a Higgs signal would be
detectable in the $\nu\bar\nu VV$ channel up to $m_H=350$~GeV.

\section{summary}

We have studied Standard-Model Higgs-boson signals at a possible future
$e^+e^-$ collider with CM energy $\sqrt s=0.5$~TeV. Our results may be
summarized as follows:

\begin{enumerate}

\item[(i)]
 For an intermediate-mass Higgs boson in the range $M_Z<m_H<2M_W$, the
$ZH$-production channel offers the biggest production cross section, even
below the nominal $ZH$ threshold; see Fig.~\ref{s-depend}.
 We have updated the branching fractions into different decay modes, which
determine the detectability, for this mass range; see Fig.~\ref{branch}.

\item[(ii)]
For a heavy Higgs boson, with $m_H>2M_W$, the most promising signals are
in the channels $e^+e^-\rightarrow \nu\bar \nu H\rightarrow \nu\bar \nu VV$,
where $V=W,Z$ and $V\rightarrow jj$ dijet decays are detected. Various
backgrounds from other $\nu\bar\nu VV$, $e\nu VV$, $eeVV$, and $t\bar t$
production mechanisms can be greatly suppressed by a $V$-rapidity cut,
a missing transverse momentum $\bigl(=p_T(VV)\bigr)$ cut and a
central-lepton veto. A detectable $\nu\bar\nu jjjj$ signal is predicted up
to masses of order $m_H=300$--350~GeV;
see Tables~\ref{various-in}, V and Figs.~\ref{103}(b), \ref{fifteen}(b).
Assuming an annual integrated luminosity of
$50\,{\rm fb}^{-1}$, 50\% instrumental efficiency, and a pessimistic
beamstrahlung scenario, a mass value $m_H=300\ (350)$~GeV
would imply about 30 (10) signal events per year in a narrow $m(jjjj)$ peak,
with about 50 (15) background events under this peak.

\item[(iii)]
Another, smaller, Higgs-boson signal appears in the channel $e^+e^-\rightarrow
ZH\rightarrow\mu^+\mu^- VV$. This is not threatened by large backgrounds and
requires no stringent cuts. For the same assumed luminosity and efficiency as
above a $\mu\mu jjjj$ signal of about 5 (3) events/year above a relatively
small background may  be expected for $m_H=250\ (300)$~GeV;
see Tables~\ref{mumu-in}, VI and Figs.~\ref{103}(c), \ref{fifteen}(c).
This signal arises entirely from the $ZH$ diagram. The cross
section for this signal is a factor of 10--40 times smaller than that in the
$\nu \bar \nu VV$ channels.

\item[(iv)]
 A third Higgs-boson signal appears in the channel $e^+e^-\to ZH\to 6\,$jets.
The signal here is comparable in size with the $\nu\bar\nu jjjj$ signal,
but the background is much larger; precise
predictions would require detailed jet simulation studies.

\item[(v)]
We have used minimal illustrative cuts to bring the backgrounds under
control.  Further suppression of the $t\bar t$ backgrounds can in principle
be achieved by vetoing $\nu\bar\nu VV$ candidate events where $m(jjj) \approx
m_t$, and $\mu\mu VV$ candidate events where one muon is not isolated.  The
$m(jjj)$ cut can however be damaging to the Higgs-boson signal, depending on
$m_
t$
and $m_H$.

\item[(vi)]
The $\nu\bar\nu H$ signal receives contributions both from $ZH$-production
and from $WW$-fusion mechanisms. These contributions can in principle
be separated on the basis of their different $p_T(VV)$ dependences in the
$\nu\bar\nu VV$ channel, or by comparing the $\nu\bar \nu VV$ and $\mu\mu VV$
signals. If these contributions are measured separately, they will allow a
direct comparison of the $ZZH$ and $WWH$ coupling strengths.

\item[(vii)]
Bremsstrahlung and beamstrahlung corrections are very important. They lower the
subprocess CM energy and momentum, enhancing annihilation channels and
suppressing scattering channels of Higgs-boson production, and smearing out
some kinematical features like Jacobian peaks in $p_T(VV)$. Our calculations of
heavy Higgs-boson signals include both initial-state bremsstrahlung effects and
illustrative calculations of beamstrahlung. The net effect in our present
cases is to reduce signals and increase backgrounds.

\item[(viii)]
The measurability of heavy Higgs-boson signals at an $e^+e^-$ collider
with $\sqrt s=0.5$~TeV is illustrated most dramatically in Figs.~\ref{103} and
15. Here all major backgrounds and all our selection cuts plus bremsstrahlung
and illustrative beamstrahlung
effects are fully included; experimental smearing in the
$H\rightarrow VV$ invariant mass distribution is still to be added.

\end{enumerate}

\acknowledgements
We thank Tim Barklow for making available to us his beamstrahlung simulation
program,
Tao Han for valuable suggestions during the initial stages of this work,
San Fu Tuan for encouraging us to complete this analysis,
and Peter Zerwas for proposing a comparative study of the beamstrahlung
performances of the various NLC designs.
This work is supported in part by the U.S.~Department of Energy under contract
No.~DE-AC02-76ER00881 and in part by the University of Wisconsin
Research Committee with funds granted by the Wisconsin Alumni Research
Foundation.

\newpage
\appendix{Feynman Amplitudes}
In this appendix we list the matrix elements for all the considered
processes, from which explicit helicity amplitudes can be directly computed.
To start with, we introduce some general notation:
\begin{eqnarray}
g_a^W(f) & = & -g_b^W(f) = \frac{g}{2 \sqrt{2}} \, , \\
g_a^Z(f) & = & g_Z \left( {T_{3f}\over2} - Q_f x_{\rm w}\right) \, ,\\
g_b^Z(f) & = & - g_Z {T_{3f}\over2} \, ,\\
g_a^\gamma(f) & = & e Q_f\, ,\\
g_b^\gamma(f) & = & 0\, ,\\
g^V(f) & = & g_a^V(f) + g_b^V(f) \gamma^5\,\qquad(V=\gamma,W,Z)\,,\\
D^X(k) & = & \frac{1}{k^2-M_X^2 + i \Gamma_X(k^2) M_X}\,,\qquad
\Gamma_X(k^2) = \Gamma_X \theta(k^2) \nonumber \\
&&  \qquad (\mbox{with }X=\gamma,W,Z,H) \, ,\\
P_V^{\alpha \beta}(k) & = &  \left [ g^{\alpha \beta} + \frac{(1-\xi)k^\alpha
k^\beta}{\xi k^2 - M_V^2} \right ] D^V(k) \,, \\
\Gamma^\alpha (k_1,k_2;\epsilon_1,\epsilon_2) & = & (k_1-k_2)^\alpha \epsilon_1
\cdot \epsilon_2 + (2k_2+k_1) \cdot \epsilon_1 \epsilon_2^\alpha
- (2k_1+k_2) \cdot \epsilon_2 \epsilon_1^\alpha\, , \\
g_{VWW} & = & \left \{
               \begin{array}{ll}
                e \cot \theta_{\rm w}  \quad & {\rm for\ } V=Z \\
                e                      & {\rm for\ } V=\gamma \, .
               \end{array} \right.  \\ \nonumber
\end{eqnarray}
Here $Q_f$ and $T_{3f}$ are the electric charge (in units of the positron
charge) and the third component of weak isospin of the fermion $f$, $g$ is
the SU(2) gauge coupling, and $g_Z=g/\cos \theta_{\rm w}$,
$x_{\rm w}=\sin^2 \theta_{\rm w}$, with $\theta_{\rm w}$ being the weak
mixing angle in the Standard Model.  Dots between 4-vectors denote scalar
products and $g_{\alpha \beta}$ is the Minkowskian metric tensor with
$g_{00}=-g_{11}=-g_{22}=-g_{33}=1$; $\xi$ is a gauge-fixing parameter.

In Fig.~\ref{e->nuW(Z)}, $p_i\ (i=1,\ldots,4)$ denote the momenta flowing along
the corresponding fermion lines in the direction of the arrows.
We shall always denote the associated spinors by the symbols $u(p_i)$ and $\bar
u(p_i)$ for the ingoing and outgoing arrows, which is usual for the
annihilation and creation of fermions, respectively. When an ingoing arrow
represents the production of an antifermion with physical momentum $p_i$
(momentum label $-p_i$), the symbol $u(-p_i)$ is defined to mean $v(p_i)$.  A
similar
convention applies to fermion (antifermion) spins, although the spin labels
are not written explicitly below. The symbols $\chi^\pm$ in
Fig.~\ref{e->nuW(Z)}(e) and (g) denote the Goldstone bosons in $R_\xi$ gauge.

Finally, when adding the contributions from annihilation and
scattering together, a relative sign change has to be introduced
 between the two sets of formulas given below. The spin labels of the
spinors $u$ and $\bar u$, and the polarization labels
of the $\epsilon$ are not shown explicitly. To obtain the final cross
section, we sum over all the final fermion (antifermion) spins and weak boson
polarizations, and average the initial fermion (antifermion) spins.

\subsection{$e^+e^- \rightarrow \nu \bar \nu W^+ W^-$ lepton scattering
contributions}

Lepton scattering comes from the generic diagrams depicted in
Figs.~\ref{e->nuW(Z)}(a)--(d), (f), (h)--(o), (p), (q), (s), (t), and (v)--(y)
with the following substitutions:
\[
\begin{array}{rlrl}
1  &\rightarrow   e^-\, , &       2  &\rightarrow \nu_e\, ,\\
3  &\rightarrow   \nu_e\, ,  &    4  &\rightarrow  e^-\, ,\\
V_1  &\rightarrow   W^-\, , & \quad  V_2  &\rightarrow  W^+\, .
\end{array}
\]
In this channel fermion lines 1 and 4 represent the incoming particles.
It is convenient to introduce the following short-hand notations :
\begin{eqnarray}
\nonumber
u_i & = & u(p_i)\;, \hspace{0.2in} \bar u_i = \bar u (p_i)\;,
\hspace{0.2in} \epsilon_i = \epsilon(k_i) \;,\\
\nonumber
J_1^\alpha & = & \bar u_2 \gamma^\alpha g^W(e) u_1 \, D^W(p_1-p_2)\, ,\\
\nonumber
J_2^\alpha & = & \bar u_4 \gamma^\alpha g^W(e) u_3 \, D^W(p_3-p_4)\, ,\\
\bar u_{ij}^{(m)} & = & \bar u_i \overlay{/}{\epsilon}_j g^W(e) \frac
{\overlay{/}{p}_i + \overlay{/}{k}_j + m}{(p_i+k_j)^2 -m^2}\, ,\\
\nonumber
u_{ji}^{(m)} & = & \frac{\overlay{/}{p}_i - \overlay{/}{k}_j + m}{(p_i-k_j)^2
-m^2}  \overlay{/}{\epsilon}_j g^W(e) u_i\, ,\\
\nonumber
\bar u_{2ij}^{(m,m')} & = & \bar u_{2i}^{(m)} \overlay{/}{\epsilon}_j g^W(e)
\frac{\overlay{/}{p}_2 + \overlay{/}{k}_i + \overlay{/}{k}_j + m'}
{(p_2+k_i+k_j)^2 -m'^2}\, ,\\
\nonumber
u_{ji1}^{(m',m)} & = &
\frac{\overlay{/}{p}_1 - \overlay{/}{k}_i - \overlay{/}{k}_j + m'}
{(p_1-k_i-k_j)^2 -m'^2} \overlay{/}{\epsilon}_j g^W(e) u_{i1}^{(m)}\, ,\\
\nonumber
\end{eqnarray}
where $m,m'$ denote the (finite) electron mass or the (zero) neutrino mass
in fermion propagators.  Then the amplitudes read
\begin{eqnarray}
\nonumber
{\cal M}^{(a)} & = & -g^2 M_W^2 D^H(k_1+k_2) J_1 \cdot J_2 \epsilon_1 \cdot
\epsilon_2\,
,\\
\nonumber
{\cal M}^{(b)} & = & -g^2 M_W^2 D^H(p_1-p_2-k_1) J_1 \cdot \epsilon_1 J_2 \cdot
\epsilon_2\, ,\\
\nonumber
{\cal M}^{(c)} & = & g^2 \left [2 J_2 \cdot \epsilon_1 J_1 \cdot \epsilon_2 -
J_2 \cdot \epsilon_2 J_1 \cdot \epsilon_1 -
J_2 \cdot J_1 \epsilon_1 \cdot \epsilon_2 \right ]\, ,\\
\nonumber
{\cal M}^{(d)} & = & \sum_{V=\gamma,Z} g_{VWW}^2 P_V^{\alpha \beta}(k_1+k_2)
\Gamma_\alpha
(p_1-p_2,p_3-p_4;J_1,J_2) \Gamma_\beta(k_1,k_2;\epsilon_1,\epsilon_2)\, ,\\
\nonumber
{\cal M}^{(f)} & = & \sum_{V=\gamma,Z} g_{VWW}^2 P_V^{\alpha
\beta}(p_1-p_2-k_1)
\Gamma_\alpha
(p_1-p_2,-k_1;J_1,\epsilon_1) \Gamma_\beta(-k_2,p_3-p_4;\epsilon_2,J_2)\, ,\\
\nonumber
{\cal M}^{(h-k)} & = &\sum_{V=\gamma,Z} P_V^{\alpha \beta}(p_1-p_2-k_1) \left [
\bar
u_{21}^{(m)} \gamma_\alpha g^V(e) u_1 + \bar u_2 \gamma_\alpha g^V(\nu)
u_{11}^{(0)}
\right ] \\ \nonumber
&& \qquad {}\times\left [\bar u_{42}^{(0)} \gamma_\beta g^V(\nu) u_3 + \bar
u_4 \gamma_\beta g^V(e)
  u_{23}^{(m)} \right ]\, ,\\
\nonumber
{\cal M}^{(l,m)} & = & \sum_{V=\gamma,Z} - g_{VWW} P_V^{\alpha
\beta}(p_1-p_2-k_1)
\Gamma_\alpha
(p_1-p_2,-k_1;J_1,\epsilon_1) \\ \nonumber
&&\qquad {}\times\left [\bar u_{42}^{(0)} \gamma_\beta g^V(\nu) u_3 + \bar
u_4
\gamma_\beta g^V(e)
u_{23}^{(m)} \right ]\, ,\\
%
{\cal M}^{(n,o)} & = & \sum_{V=\gamma,Z} - g_{VWW} P_V^{\alpha
\beta}(p_1-p_2-k_1)
\Gamma_\alpha
(-k_2,p_3-p_4;\epsilon_2,J_2) \\ \nonumber
&&\qquad {}\times\left [\bar u_{21}^{(m)} \gamma_\beta g^V(e) u_1 + \bar u_2
\gamma_\beta g^V(\nu)
u_{11}^{(0)} \right ]\, ,\\
\nonumber
{\cal M}^{(p,q)} & = & \bar u_{212}^{(m,0)} \overlay{/}{J}_2 g^W(e) u_1 + \bar
u_2
\overlay{/}{J}_2 g^W(e) u_{211}^{(m,0)}\, ,\\
\nonumber
{\cal M}^{(s,t)} & = & \bar u_{421}^{(0,m)} \overlay{/}{J}_1 g^W(e) u_3 + \bar
u_4
\overlay{/}{J}_1 g^W(e) u_{123}^{(0,m)}\, ,\\
\nonumber
{\cal M}^{(v)} & = & \sum_{V=\gamma,Z} - g_{VWW} D^V(k_1+k_2) \bar u_4
\overlay{/}{\Gamma}(k_1,k_2;\epsilon_1,\epsilon_2) g^V(e)
\frac{\overlay{/}{p}_4 + \overlay{/}{k}_1 + \overlay{/}{k}_2 +m}
{(p_4+k_1+k_2)^2 - m^2} \overlay{/}{J}_1 g^W(e) u_3 \, ,\\
\nonumber
{\cal M}^{(w)} & = & - g_{ZWW} D^Z(k_1+k_2) \bar u_4 \overlay{/}{J}_1 g^W(e)
\frac
{\overlay{/}{p}_3 - \overlay{/}{k}_1 - \overlay{/}{k}_2}
{(p_3-k_1-k_2)^2} \overlay{/}{\Gamma}(k_1,k_2;\epsilon_1,\epsilon_2) g^Z(\nu)
u_3\, ,\\
\nonumber
{\cal M}^{(x)} & = & \sum_{V=\gamma,Z} - g_{VWW} D^V(k_1+k_2) \bar u_2
\overlay{/}{J}_2
g^W(e) \frac{\overlay{/}{p}_1 - \overlay{/}{k}_1 - \overlay{/}{k}_2 +m}
{(p_1-k_1-k_2)^2 -m^2} \overlay{/}{\Gamma}(k_1,k_2;\epsilon_1,\epsilon_2)
g^V(e) u_1\, ,\\
\nonumber
{\cal M}^{(y)} & = & -g_{ZWW} D^Z(k_1+k_2) \bar u_2
\overlay{/}{\Gamma}(k_1,k_2;\epsilon_1,\epsilon_2) g^Z(\nu) \frac
{\overlay{/}{p}_2 + \overlay{/}{k}_1 + \overlay{/}{k}_2 }
{(p_2+k_1+k_2)^2} \overlay{/}{J}_2 g^W(e) u_1 \, .
\end{eqnarray}
%
\subsection{$e^+e^- \rightarrow \nu \bar \nu W^+ W^-$ lepton annihilation
contributions}

Lepton annihilation contributes via the generic diagrams depicted in
Figs.~\ref{e->nuW(Z)}(a), (c), (f), (g), (h), (j), and (l)--(y)
with the following substitutions:
\[
\begin{array}{rlrl}
1  &\rightarrow   e^-\, , &\quad  2  &\rightarrow e^-\, ,\\
3  &\rightarrow   \nu\, ,  &\quad 4  &\rightarrow  \nu\, ,\\
V_1  &\rightarrow   W^-\, , &\quad V_2  &\rightarrow  W^+\, .
\end{array}
\]
Here fermion lines 1 and 2 represent the incoming particles.
Redefining
\begin{eqnarray}
\nonumber
J_{1 \alpha}^V & = & \bar u_2 \gamma_\alpha g^V(e) u_1 D^V(p_1-p_2)\, ,\\
J_{2 \alpha} & = & \bar u_4 \gamma_\alpha g^Z(\nu) u_3 D^Z(p_3-p_4)\, ,
\end{eqnarray}
we find the amplitudes:
\begin{eqnarray}
\nonumber
{\cal M}^{(a)} & = & - \frac{g^2}{1-x_{\rm w}} M_W^2 D^H(k_1+k_2) \, \epsilon_1
\cdot
\epsilon_2
J_1^Z \cdot J_2\, ,\\
\nonumber
{\cal M}^{(c)} & = & \sum_{V=\gamma,Z} - g_{VWW} g_{ZWW} \left [
2 \epsilon_1 \cdot \epsilon_2 J_1^V \cdot J_2 -
\epsilon_1 \cdot J_1^V \epsilon_2 \cdot J_2 -
\epsilon_1 \cdot J_2 \epsilon_2 \cdot J_1^V \right ]\, ,\\
\nonumber
{\cal M}^{(f)} & = & \sum_{V=\gamma,Z} \left[ g_{VWW} g_{ZWW} P_W^{\alpha
\beta}(p_1-p_2-k_1)
\Gamma_\alpha(-k_1,p_1-p_2;\epsilon_1,J_1^V)
\Gamma_\beta(p_3-p_4,-k_2;J_2,\epsilon_2)\, \right. \\ \nonumber
 &  &  \left. \quad{}+  g_{VWW} g_{ZWW} P_W^{\alpha \beta}(p_1-p_2-k_2)
\, \Gamma_\alpha(p_1-p_2,-k_2;J_1^V,\epsilon_2)
\, \Gamma_\beta(-k_1,p_3-p_4;\epsilon_1,J_2) \right] \, ,\\
\nonumber
{\cal M}^{(g)} & = & \sum_{V=\gamma,Z} g^2 x_{\rm w} M_W^2 J_1^V \cdot
\epsilon_1 \, J_2
\cdot
\epsilon_2 \, \frac{\xi}{\xi (p_1-p_2-k_1)^2 -M_W^2} \\ \nonumber
& & \qquad \times \left \{  \begin{array}{ll}
           - \tan^2\theta_{\rm w}\,,  & \mbox{  if } V=Z \\
           \phantom-  \tan\theta_{\rm w}\,,    & \mbox{  if } V=\gamma
            \end{array}  \right.
 \;\; \;\; {}+ (k_1 \leftrightarrow k_2)\, ,\\
\nonumber
{\cal M}^{(h)} & = & P_W^{\alpha \beta}(p_1-p_2-k_2) \, \bar u_{22}^{(0)}
\gamma_\alpha
g^W(e) u_1
\bar u_{41}^{(m)} \gamma_\beta g^W(e) u_3 \, ,\\
\nonumber
{\cal M}^{(j)} & = & P_W^{\alpha \beta}(p_1-p_2-k_1) \, \bar u_2 \gamma_\alpha
g^W(e)
u_{11}^{(0)}
\bar u_4 \gamma_\beta g^W(e) u_{23}^{(m)} \, ,\\
%
{\cal M}^{(l)} & = & \sum_{V=\gamma,Z} - g_{VWW} P_W^{\alpha
\beta}(p_1-p_2-k_2)
\Gamma_\alpha
(p_1-p_2,-k_2;J_1^V,\epsilon_2) \, \bar u_{41}^{(m)} \gamma_\beta g^W(e) u_3
\, ,\\
\nonumber
{\cal M}^{(m)} & = & \sum_{V=\gamma,Z} - g_{VWW} P_W^{\alpha
\beta}(p_1-p_2-k_1)
\Gamma_\alpha
(-k_1,p_1-p_2;\epsilon_1,J_1^V) \, \bar u_4 \gamma_\beta g^W(e) u_{23}^{(m)}
\, ,\\
\nonumber
{\cal M}^{(n)} & = & - g_{ZWW} P_W^{\alpha \beta}(p_1-p_2-k_2) \Gamma_\alpha
(-k_1,p_3-p_4;\epsilon_1,J_2) \, \bar u_{22}^{(0)} \gamma_\beta g^W(e) u_1
\, ,\\
\nonumber
{\cal M}^{(o)} & = & - g_{ZWW} P_W^{\alpha \beta}(p_1-p_2-k_1) \Gamma_\alpha
(p_3-p_4,-k_2;J_2,\epsilon_2) \, \bar u_2 \gamma_\beta g^W(e) u_{11}^{(0)}
\, ,\\
\nonumber
{\cal M}^{(p-r)} & = & \bar u_{22}^{(0)} \overlay{/}{J}_2 g^Z(\nu) u_{11}^{(0)}
+
\bar {u}_2 \overlay{/}{J}_2 g^Z(e) u_{211}^{(m,0)}
+ \bar {u}_{221}^{(0,m)} \overlay{/}{J}_2 g^Z(e) u_1 \, ,\\
\nonumber
{\cal M}^{(s-u)} & = & \bar u_4 \overlay{/}{J}_1^Z g^Z(\nu) u_{123}^{(0,m)} +
\sum_{V=\gamma,Z} \bar {u}_{41}^{(m)} \overlay{/}{J}_1^V g^V(e) u_{23}^{(m)}
+ \bar {u}_{412}^{(m,0)} \overlay{/}{J}_1^Z g^Z(\nu) u_3 \, ,\\
\nonumber
{\cal M}^{(v)} & = & -g_{ZWW} D^Z(k_1+k_2) \, \bar u_4
\overlay{/}{\Gamma}(k_1,k_2;\epsilon_1,\epsilon_2) g^Z(\nu) \frac{\overlay{/}
{p}_4 + \overlay{/}{k}_1 + \overlay{/}{k}_2}
{(p_4+k_1+k_2)^2 } \overlay{/}{J}_1^Z g^Z(\nu) u_3 \, ,\\
\nonumber
{\cal M}^{(w)} & = & -g_{ZWW} D^Z(k_1+k_2) \, \bar u_4
\overlay{/}{J}_1^Z g^Z(\nu) \frac{\overlay{/}
{p}_3 - \overlay{/}{k}_1 - \overlay{/}{k}_2 }
{(p_3-k_1-k_2)^2 } \overlay{/}{\Gamma}(k_1,k_2;\epsilon_1,\epsilon_2)
 g^Z(\nu) u_3 \, ,\\
\nonumber
{\cal M}^{(x)} & = & \sum_{V=\gamma,Z} -g_{VWW} D^V(k_1+k_2) \, \bar u_2
\overlay{/}{J}_2 g^Z(e) \frac{\overlay{/}
{p}_1 - \overlay{/}{k}_1 - \overlay{/}{k}_2 +m}
{(p_1-k_1-k_2)^2 -m^2} \overlay{/}{\Gamma}(k_1,k_2;\epsilon_1,\epsilon_2)
 g^V(e) u_1 \, ,\\
\nonumber
{\cal M}^{(y)} & = & \sum_{V=\gamma,Z} -g_{VWW} D^V(k_1+k_2) \,\bar u_2
\overlay{/}{\Gamma}(k_1,k_2;\epsilon_1,\epsilon_2) g^V(e) \frac{\overlay{/}
{p}_2 + \overlay{/}{k}_1 + \overlay{/}{k}_2 +m}
{(p_2+k_1+k_2)^2 -m^2} \overlay{/}{J}_2 g^Z(e) u_1 \, .
\end{eqnarray}
%
\subsection{$e^+e^- \rightarrow \nu \bar \nu Z Z$ lepton scattering
contributions}
Lepton scattering comes from the generic diagrams depicted in
Figs.~\ref{e->nuW(Z)}(a), (c), and (f)--(u)
with the following substitutions:
\[
\begin{array}{rl@{\quad}rl}
1  &\rightarrow   e^-\, , &  2  &\rightarrow \nu_e\, ,\\
3  &\rightarrow   \nu_e\, ,  & 4  &\rightarrow  e^-\, ,\\
V_1 & \rightarrow   Z\, , & V_2  &\rightarrow  Z\, .
\end{array}
\]
Here fermion lines 1 and 4 represent the incoming particles.
Redefining
\begin{eqnarray}
\nonumber
J_{1 \alpha} & = & \bar u_2 \gamma_\alpha g^W(e) u_1 \, D^W(p_1-p_2)\, ,\\
\nonumber
J_{2 \alpha} & = & \bar u_4 \gamma_\alpha g^W(e) u_3 \, D^W(p_3-p_4)\, ,\\
\nonumber
u_{ji,f}^{(m)} & = & \frac{\overlay{/}{p}_i -\overlay{/}{k}_j +m}
{(p_i-k_j)^2 - m^2} \overlay{/}{\epsilon}_j g^Z(f) u_i \, ,\\
\bar u_{ij,f}^{(m)} & = & \bar u_i\overlay{/}{\epsilon}_j g^Z(f)
\frac{\overlay{/}{p}_i +\overlay{/}{k}_j +m}
{(p_i+k_j)^2 - m^2} \, ,\\
\nonumber
\bar u_{2ij,f_1f_2}^{(m,m')} & = & \bar u_{2i,f_1}^{(m)}
\overlay{/}{\epsilon}_j
g^Z(f_2)\frac{\overlay{/}{p}_2 +\overlay{/}{k}_i + \overlay{/}{k}_j +m'}
{(p_2+k_i+k_j)^2 - m'^2} \, ,\\
\nonumber
u_{ji1,f_2f_1}^{(m',m)} & = &\frac{\overlay{/}{p}_1 - \overlay{/}{k}_i  -
\overlay{/}{k}_j +m'} {(p_1- k_i- k_j)^2 - m'^2}
 \overlay{/}{\epsilon}_j
g^Z(f_2) u_{i1,f_1}^{(m)} \, ,
\end{eqnarray}
where $f,f_1$ and $f_2$ in the subscript denote the fermions to which the
external $Z$-bosons are attached.  We find the amplitudes:
\begin{eqnarray}
\nonumber
{\cal M}^{(a)} & = & - \frac{g^2}{1-x_{\rm w}} M_W^2 D^H(k_1+k_2) \, \epsilon_1
\cdot
\epsilon_2 \, J_1 \cdot J_2 \; ,\\
\nonumber
{\cal M}^{(c)} & = & - g^2 (1-x_{\rm w})  \left[ 2 \epsilon_1 \cdot \epsilon_2
\, J_1 \cdot
J_2 - \epsilon_1 \cdot J_1 \, \epsilon_2 \cdot J_2 -
\epsilon_1 \cdot J_2 \, \epsilon_2 \cdot J_1 \right] \;, \\
\nonumber
{\cal M}^{(f)} & = & g_{ZWW}^2 P_W^{\alpha \beta}(p_1-p_2-k_1)
\Gamma_\alpha (-k_1,p_1-p_2;\epsilon_1,J_1) \Gamma_\beta
(p_3-p_4,-k_2;J_2,\epsilon_2)
\; ,\\
\nonumber
{\cal M}^{(g)} & = & - \frac{g^2}{1-x_{\rm w}} x_{\rm w}^2 M_W^2 \epsilon_1
\cdot J_1 \,
\epsilon_2 \cdot J_2 \, \frac{\xi}{\xi(p_1-p_2-k_1)^2 -M_W^2}\; ,\\
\nonumber
{\cal M}^{(h-k)} & = & P_W^{\alpha \beta}(p_1-p_2-k_1) \left[ \bar u_2
\gamma_\alpha g^W(e)
u_{11,e}^{(m)}  +  \bar u_{21,\nu}^{(0)} \gamma_\alpha g^W(e) u_1 \right]
\\ \nonumber
&& \quad \times \left[\bar u_4 \gamma_\beta g^W(e) u_{23,\nu}^{(0)} +
\bar u_{42,e}^{(m)} \gamma_\beta g^W(e) u_3 \right] \; ,\\
%
{\cal M}^{(l,m)} & = & -g_{ZWW} P_W^{\alpha \beta}(p_1-p_2-k_1)
\Gamma_\alpha(-k_1,p_1-p_2;\epsilon_1,J_1) \\ \nonumber
&& \quad \times \left[\bar u_4 \gamma_\beta g^W(e) u_{23,\nu}^{(0)} +
\bar u_{42,e}^{(m)} \gamma_\beta g^W(e) u_3 \right] \; ,\\
\nonumber
{\cal M}^{(n,o)} & = & -g_{ZWW} P_W^{\alpha \beta}(p_1-p_2-k_1)
\Gamma_\alpha(p_3-p_4,-k_2;J_2,\epsilon_2) \\ \nonumber
&& \quad \times \left[ \bar u_2 \gamma_\beta g^W(e)
u_{11,e}^{(m)}  +  \bar u_{21,\nu}^{(0)} \gamma_\beta g^W(e) u_1 \right]\;,\\
\nonumber
{\cal M}^{(p-r)} & = & \bar u_{22,\nu}^{(0)} \overlay{/}{J}_2 g^W(e)
u_{11,e}^{(m)} +
\bar u_{221,\nu\nu}^{(0,0)} \overlay{/}{J}_2 g^W(e) u_1 +
\bar u_2 \overlay{/}{J}_2 g^W(e) u_{211,ee}^{(m,m)} \;,\\
\nonumber
{\cal M}^{(s-u)} & = & \bar u_{42,e}^{(m)} \overlay{/}{J}_1 g^W(e)
u_{13,\nu}^{(0)} +
\bar u_{421,ee}^{(m,m)} \overlay{/}{J}_1 g^W(e) u_3 +
\bar u_4 \overlay{/}{J}_1 g^W(e) u_{213,\nu\nu}^{(0,0)} \;.
\end{eqnarray}
In the case of ${\cal M}^{(f-u)}$ it is understood that similar contributions
with
$(k_1\leftrightarrow k_2,\ \epsilon_1\leftrightarrow\epsilon_2)$ are to be
added due to identical $Z$ bosons
in the final state.
%
\subsection{$e^+e^- \rightarrow \nu \bar \nu ZZ$ lepton annihilation
contributions}
Lepton annihilation contributes via the generic diagrams depicted in
Figs.~\ref{e->nuW(Z)}(a), (b), (h)--(k), and (p)--(u)
with the following substitutions:
\[
\begin{array}{rlrl}
1  &\rightarrow   e^-\, , &\quad  2  &\rightarrow e^-\, ,\\
3  &\rightarrow   \nu\, ,  &\quad 4  &\rightarrow  \nu\, ,\\
V_1  &\rightarrow   Z\, , &\quad V_2  &\rightarrow  Z\, .
\end{array}
\]
Here fermion lines 1 and 2 represent the incoming particles.
Redefining
\begin{eqnarray}
\nonumber
J_{1 \alpha} & = & \bar u_2 \gamma_\alpha g^Z(e) u_1 D^Z(p_1-p_2)\;,\\
J_{2 \alpha} & = & \bar u_4 \gamma_\alpha g^Z(\nu) u_3 D^Z(p_3-p_4)\;,
\end{eqnarray}
we find the amplitudes:
\begin{eqnarray}
\nonumber
{\cal M}^{(a)} & = & -\frac{g^2}{(1-x_{\rm w})^2}M_W^2 D^H(k_1+k_2) \, J_1
\cdot J_2 \,
\epsilon_1 \cdot \epsilon_2 \;,\\
\nonumber
{\cal M}^{(b)} & = & -\frac{g^2}{(1-x_{\rm w})^2} M_W^2 D^H(p_1-p_2-k_1) \,
\epsilon_1
\cdot J_1 \,
\epsilon_2 \cdot J_2 \;,\\
%
{\cal M}^{(h-k)} & = & P_Z^{\alpha \beta}(p_1-p_2-k_1) \left[ \bar u_2
\gamma_\alpha g^Z(e)
u_{11,e}^{(m)} + \bar u_{21,e}^{(m)} \gamma_\alpha g^Z(e) u_1 \right] \\
\nonumber
&& \quad \times \left[\bar u_4 \gamma_\beta g^Z(\nu) u_{23,\nu}^{(0)} +
\bar u_{42,\nu}^{(0)} \gamma_\beta g^Z(\nu) u_3 \right] \;,\\
\nonumber
{\cal M}^{(p-r)} & = & \bar u_{22,e}^{(m)} \overlay{/}{J}_2 g^Z(e)
u_{11,e}^{(m)} +
\bar u_2 \overlay{/}{J}_2 g^Z(e) u_{211,ee}^{(m,m)} +
\bar u_{221,ee}^{(m,m)} \overlay{/}{J}_2 g^Z(e) u_1 \;,\\
\nonumber
{\cal M}^{(s-u)} & = & \bar u_{42,\nu}^{(0)} \overlay{/}{J}_1 g^Z(\nu)
u_{13,\nu}^{(0)} +
\bar u_4 \overlay{/}{J}_1 g^Z(\nu) u_{213,\nu\nu}^{(0,0)} +
\bar u_{421,\nu\nu}^{(0,0)} \overlay{/}{J}_1 g^Z(\nu) u_3 \;.
\end{eqnarray}
For the above amplitudes, except for ${\cal M}^{(a)}$, similar contributions
with $(k_1 \leftrightarrow k_2,\ \epsilon_1\leftrightarrow\epsilon_2)$ have to
be added due to identical $Z$
bosons in the final state.
%
%
\subsection{$e^+e^- \rightarrow e^+ e^- W^+ W^-$
: scattering and annihilation contributions}
Both scattering and annihilation
contributions arise from the generic diagrams depicted in
Figs.~\ref{e->nuW(Z)}(a), (c), (f), (g), (i), (k), and (l)--(y)
with the following substitutions:
\[
\begin{array}{rlrl}
1  &\rightarrow   e^-\, , &\quad  2  &\rightarrow e^-\, ,\\
3  &\rightarrow   e^-\, ,  &\quad 4  &\rightarrow  e^-\, ,\\
V_1  &\rightarrow   W^-\, , &\quad V_2  &\rightarrow  W^+\, .
\end{array}
\]
Here fermion lines 1 and 4 (1 and 2) are taken to represent the
incoming particles in the case of scattering (annihilation).
To obtain the full answer, both sets of diagrams have to be
added with a relative minus sign between them.
Redefining
\begin{eqnarray}
\nonumber
J_{1 \alpha}^{V_1} & = & \bar u_2 \gamma_\alpha g^{V_1}(e) u_1
D^{V_1}(p_1-p_2)\, ,\\
J_{2 \alpha}^{V_2} & = & \bar u_4 \gamma_\alpha g^{V_2}(e) u_3
D^{V_2}(p_3-p_4)\, ,
\end{eqnarray}
we find the amplitudes:
\begin{eqnarray}
\nonumber
{\cal M}^{(a)} & = & - \frac{g^2}{1-x_{\rm w}}M_W^2 D^H(k_1+k_2) \, \epsilon_1
\cdot
\epsilon_2
J_1^Z \cdot J_2^Z\, ,\\
\nonumber
{\cal M}^{(c)} & = & \sum_{V_1,V_2=\gamma,Z} - g_{V_1WW} g_{V_2WW} \left [
2 \epsilon_1 \cdot \epsilon_2 J_1^{V_1} \cdot J_2^{V_2} -
\epsilon_1 \cdot J_1^{V_1} \epsilon_2 \cdot J_2^{V_2} -
\epsilon_1 \cdot J_2^{V_2} \epsilon_2 \cdot J_1^{V_1} \right ]\, ,\\
\nonumber
{\cal M}^{(f)} & = & \sum_{V_1,V_2=\gamma,Z} \left[ g_{V_1WW} g_{V_2WW}
P_W^{\alpha
\beta}(p_1-p_2-k_1)
\Gamma_\alpha(-k_1,p_1-p_2;\epsilon_1,J_1^{V_1})
\Gamma_\beta(p_3-p_4,-k_2;J_2^{V_2},\epsilon_2)\, \right. \\ \nonumber
 &  & \qquad \left. +  g_{V_1WW} g_{V_2WW} P_W^{\alpha \beta}(p_1-p_2-k_2)
\, \Gamma_\alpha(p_1-p_2,-k_2;J_1^{V_1},\epsilon_2)
\, \Gamma_\beta(-k_1,p_3-p_4;\epsilon_1,J_2^{V_2}) \right] \, ,\\
\nonumber
{\cal M}^{(g)} & = & \sum_{V_1,V_2=\gamma,Z} g^2 x_{\rm w} M_W^2 J_1^{V_1}
\cdot \epsilon_1
\, J_2^{V_2} \cdot
\epsilon_2 \, \frac{\xi}{\xi (p_1-p_2-k_1)^2 -M_W^2} \\ \nonumber
& &  \qquad \quad \times \left \{  \begin{array}{ll}
           - \tan^2\theta_{\rm w}\,,  & \mbox{  if } V_1=V_2=Z \\
           \mbox{} -1\,,        & \mbox{  if } V_1=V_2=\gamma\\
           \mbox{}  \tan\theta_{\rm w}\,,    & \mbox{  otherwise}
            \end{array}  \right.
 \;\; \;\; {} + (k_1 \leftrightarrow k_2)\, ,\\
\nonumber
{\cal M}^{(i)} & = & P_W^{\alpha \beta}(p_1-p_2-k_1) \, \bar u_2 \gamma_\alpha
g^W(e)
u_{11}^{(0)}
\bar u_{42}^{(0)} \gamma_\beta g^W(e) u_3 \, ,\\
\nonumber
{\cal M}^{(k)} & = & P_W^{\alpha \beta}(p_1-p_2-k_2) \, \bar u_{22}^{(0)}
\gamma_\alpha
g^W(e) u_1
\bar u_4 \gamma_\beta g^W(e) u_{13}^{(0)} \, ,\\
%
{\cal M}^{(l)} & = & \sum_{V_1=\gamma,Z} - g_{V_1WW} P_W^{\alpha
\beta}(p_1-p_2-k_1)
\Gamma_\alpha
(-k_1,p_1-p_2;\epsilon_1,J_1^{V_1}) \, \bar u_{42}^{(0)} \gamma_\beta g^W(e)
u_3
\, ,\\
\nonumber
{\cal M}^{(m)} & = & \sum_{V_1=\gamma,Z} - g_{V_1WW} P_W^{\alpha
\beta}(p_1-p_2-k_2)
\Gamma_\alpha
(p_1-p_2,-k_2;J_1^{V_1},\epsilon_2) \, \bar u_4 \gamma_\beta g^W(e)
u_{13}^{(0)}
\, ,\\
\nonumber
{\cal M}^{(n)} & = & \sum_{V_2=\gamma,Z} - g_{V_2WW} P_W^{\alpha
\beta}(p_1-p_2-k_2)
\Gamma_\alpha
(-k_1,p_3-p_4;\epsilon_1,J_2^{V_2}) \, \bar u_{22}^{(0)} \gamma_\beta g^W(e)
u_1
\, ,\\
\nonumber
{\cal M}^{(o)} & = & \sum_{V_2=\gamma,Z} - g_{V_2WW} P_W^{\alpha
\beta}(p_1-p_2-k_1)
\Gamma_\alpha
(p_3-p_4,-k_2;J_2^{V_2},\epsilon_2) \, \bar u_2 \gamma_\beta g^W(e)
u_{11}^{(0)}
\, ,\\
\nonumber
{\cal M}^{(p-r)} & = & \sum_{V_2=\gamma,Z} \left[ \bar u_{22}^{(0)}
\overlay{/}{J}_2^{V_2}
 g^{V_2}(\nu) u_{11}^{(0)} +
\bar {u}_2 \overlay{/}{J}_2^{V_2} g^{V_2}(e) u_{211}^{(m,0)}
+ \bar {u}_{221}^{(0,m)} \overlay{/}{J}_2^{V_2} g^{V_2}(e) u_1 \right]\, ,\\
\nonumber
{\cal M}^{(s-u)} & = & \sum_{V_1=\gamma,Z} \left[\bar u_{42}^{(0)}
\overlay{/}{J}_1^{V_1}
 g^{V_1}(\nu) u_{13}^{(0)} +
\bar u_4 \overlay{/}{J}_1^{V_1} g^{V_1}(e) u_{213}^{(m,0)}
+ \bar {u}_{421}^{(0,m)} \overlay{/}{J}_1^{V_1} g^{V_1}(e) u_3 \right]\, ,\\
\nonumber
{\cal M}^{(v)} & = & \sum_{V_1,V_2=\gamma,Z} -g_{V_2WW} D^{V_2}(k_1+k_2) \,
\bar u_4
\overlay{/}{\Gamma}(k_1,k_2;\epsilon_1,\epsilon_2) g^{V_2}(e) \frac{\overlay{/}
{p}_4 + \overlay{/}{k}_1 + \overlay{/}{k}_2 + m}
{(p_4+k_1+k_2)^2 -m^2} \overlay{/}{J}_1^{V_1} g^{V_1}(e) u_3 \, ,\\
\nonumber
{\cal M}^{(w)} & = & \sum_{V_1,V_2=\gamma,Z} -g_{V_2WW} D^{V_2}(k_1+k_2) \,
\bar u_4
\overlay{/}{J}_1^{V_1} g^{V_1}(e) \frac{\overlay{/}
{p}_3 - \overlay{/}{k}_1 - \overlay{/}{k}_2 + m}
{(p_3-k_1-k_2)^2 - m^2} \overlay{/}{\Gamma}(k_1,k_2;\epsilon_1,\epsilon_2)
 g^{V_2}(e) u_3 \, ,\\
\nonumber
{\cal M}^{(x)} & = & \sum_{V_1,V_2=\gamma,Z} -g_{V_1WW} D^{V_1}(k_1+k_2) \,
\bar u_2
\overlay{/}{J}_2^{V_2} g^{V_2}(e) \frac{\overlay{/}
{p}_1 - \overlay{/}{k}_1 - \overlay{/}{k}_2 +m}
{(p_1-k_1-k_2)^2 -m^2} \overlay{/}{\Gamma}(k_1,k_2;\epsilon_1,\epsilon_2)
 g^{V_1}(e) u_1 \, ,\\
\nonumber
{\cal M}^{(y)} & = & \sum_{V_1,V_2=\gamma,Z} -g_{V_1WW} D^{V_1}(k_1+k_2) \,\bar
u_2
\overlay{/}{\Gamma}(k_1,k_2;\epsilon_1,\epsilon_2) g^{V_1}(e) \frac{\overlay{/}
{p}_2 + \overlay{/}{k}_1 + \overlay{/}{k}_2 +m}
{(p_2+k_1+k_2)^2 -m^2} \overlay{/}{J}_2^{V_2} g^{V_2}(e) u_1 \, .
\end{eqnarray}
%
\subsection{$e^+e^- \rightarrow e^+ e^- ZZ$
: scattering and annihilation contributions}
Both scattering and annihilation contributions arise from the generic
diagrams depicted in Figs.~\ref{e->nuW(Z)}(a), (b), (h)--(k), and (p)--(u)
with the following substitutions:
\[
\begin{array}{rlrl}
1  &\rightarrow   e^-\, , &\quad  2  &\rightarrow e^-\, ,\\
3  &\rightarrow   e^-\, ,  &\quad 4  &\rightarrow  e^-\, ,\\
V_1  &\rightarrow   Z\, , &\quad V_2  &\rightarrow  Z\, .
\end{array}
\]
Here fermion lines 1 and 4 (1 and 2) are taken to represent the
incoming particles in the case of scattering (annihilation).
To obtain the full answer, both sets of diagrams have to be
added with a relative minus sign between them.
Redefining
\begin{eqnarray}
\nonumber
J_{1 \alpha}^V & = & \bar u_2 \gamma_\alpha g^V(e) u_1 D^V(p_1-p_2)\;,\\
J_{2 \alpha}^V & = & \bar u_4 \gamma_\alpha g^V(e) u_3 D^V(p_3-p_4)\;,
\end{eqnarray}
we find the amplitudes:
\begin{eqnarray}
\nonumber
{\cal M}^{(a)} & = & -\frac{g^2}{(1-x_{\rm w})^2} M_W^2 D^H(k_1+k_2) \, J_1^Z
\cdot J_2^Z
\,
\epsilon_1 \cdot \epsilon_2 \;,\\
\nonumber
{\cal M}^{(b)} & = & -\frac{g^2}{(1-x_{\rm w})^2} M_W^2 D^H(p_1-p_2-k_1) \,
\epsilon_1
\cdot J_1^Z \,
\epsilon_2 \cdot J_2^Z \;,\\
%
{\cal M}^{(h-k)} & = & \sum_{V=\gamma,Z} P_V^{\alpha \beta}(p_1-p_2-k_1) \left[
\bar u_2
\gamma_\alpha g^V(e)
u_{11,e}^{(m)} + \bar u_{21,e}^{(m)} \gamma_\alpha g^V(e) u_1 \right] \\
\nonumber
&& \qquad \quad \times \left[\bar u_4 \gamma_\beta g^V(e) u_{23,e}^{(m)} +
\bar u_{42,e}^{(m)} \gamma_\beta g^V(e) u_3 \right] \;,\\
\nonumber
{\cal M}^{(p-r)} & = & \sum_{V=\gamma,Z} \left[ \bar u_{22,e}^{(m)}
\overlay{/}{J}_2^V
g^V(e)
 u_{11,e}^{(m)} +
\bar u_2 \overlay{/}{J}_2^V g^V(e) u_{211,ee}^{(m,m)} +
\bar u_{221,ee}^{(m,m)} \overlay{/}{J}_2^V g^V(e) u_1 \right] \;,\\
\nonumber
{\cal M}^{(s-u)} & = & \sum_{V=\gamma,Z} \left[ \bar u_{42,e}^{(m)}
\overlay{/}{J}_1^V
 g^V(e) u_{13,e}^{(m)} +
\bar u_4 \overlay{/}{J}_1^V g^V(e) u_{213,ee}^{(m,m)} +
\bar u_{421,ee}^{(m,m)} \overlay{/}{J}_1^V g^V(e) u_3 \right] \;.
\end{eqnarray}
For the above amplitudes, except for ${\cal M}^{(a)}$, similar contributions
with $(k_1 \leftrightarrow k_2,\, \epsilon_1 \leftrightarrow \epsilon_2)$
have to be added due to identical $Z$ bosons in the final state.
\subsection{$e^+ e^- \rightarrow e^+ \nu W^- Z$ : scattering and
annihilation channels}
Both scattering and annihilation contributions arise from the generic
diagrams depicted in
Figs.~\ref{e->nuW(Z)}(b)--(r), (t), (u), and (w)--(y)
with the following substitutions:
\[
\begin{array}{rlrl}
1  &\rightarrow   e^-\, , &\quad  2  &\rightarrow \nu\, ,\\
3  &\rightarrow   e^-\, ,  &\quad 4  &\rightarrow  e^-\, ,\\
V_1 & \rightarrow   W^-\, , &\quad V_2  &\rightarrow  Z\, .
\end{array}
\]
Here fermion lines 1 and 4 (3 and 4) are taken to represent the
incoming particles in the case of scattering (annihilation).
To obtain the full answer, both sets of diagrams have to be
added with a relative minus sign between them.
Redefining
\begin{eqnarray}
\nonumber
J_{1 \alpha} & = & \bar u_2 \gamma_\alpha g^W(e) u_1 D^W(p_1-p_2)\, ,\\
J_{2 \alpha}^V & = & \bar u_4 \gamma_\alpha g^V(e) u_3 D^V(p_3-p_4)\, ,
\end{eqnarray}
we find for the amplitudes:
\begin{eqnarray}
\nonumber
{\cal M}^{(b)} & = & - \frac{g^2}{1-x_{\rm w}} M_W^2 D^H(p_1-p_2-k_1)
\epsilon_1 \cdot J_1
\epsilon_2 \cdot J_2^Z \;,\\
\nonumber
{\cal M}^{(c)} & = & \sum_{V=\gamma,Z} -g_{VWW} g_{ZWW} \left[
2 \epsilon_1 \cdot J_1 \, \epsilon_2 \cdot J_2^V -
\epsilon_1 \cdot J_2^V \, \epsilon_2 \cdot J_1 -
\epsilon_1 \cdot \epsilon_2 \, J_1 \cdot J_2^V \right] \;,\\
\nonumber
{\cal M}^{(d)} & = & \sum_{V=\gamma,Z} g_{VWW} g_{ZWW} P_W^{\alpha
\beta}(k_1+k_2)
\Gamma_\alpha(k_2,k_1;\epsilon_2,\epsilon_1)
\Gamma_\beta(p_3-p_4,p_1-p_2;J_2^V,J_1) \;,\\
\nonumber
{\cal M}^{(e)} & = & \sum_{V=\gamma,Z} g^2 x_{\rm w} M_W^2 \, \epsilon_1 \cdot
\epsilon_2
J_1 \cdot J_2^{V} \; \frac{\xi}{\xi (k_1+k_2)^2 -M_W^2} \\ \nonumber
& & \qquad {} \times \left \{ \begin{array}{ll}
                          - \tan^2 \theta_{\rm w}\,,   &  \mbox{ if } V=Z \\
                          \phantom- \tan \theta_{\rm w}\,,  &\mbox{ if
}V=\gamma\;,
                            \end{array}
\right. \\
\nonumber
{\cal M}^{(f)} & = & \sum_{V=\gamma,Z} g_{VWW} g_{ZWW} P_W^{\alpha
\beta}(p_1-p_2-k_2)
\Gamma_\alpha(-k_2,p_1-p_2;\epsilon_2,J_1)
\Gamma_\beta(-k_1,p_3-p_4;\epsilon_1,J_2^V) \;,\\
\nonumber
{\cal M}^{(g)} & = & \sum_{V=\gamma,Z} g^2 x_{\rm w} M_W^2 \, \epsilon_1 \cdot
J_2^V
\epsilon_2 \cdot J_1 \; \frac{\xi}{\xi (p_1-p_2-k_2)^2 -M_W^2} \\ \nonumber
& & \qquad {} \times \left \{ \begin{array}{ll}
                          - \tan^2 \theta_{\rm w}\,,   &  \mbox{ if } V=Z \\
                          \phantom- \tan \theta_{\rm w}\,,  &\mbox{ if
}V=\gamma\;,
                        \end{array}
\right. \\
%
{\cal M}^{(h-k)} & = & P_Z^{\alpha \beta}(p_1-p_2-k_1) \, \bar u_2
\gamma_\alpha g^Z(\nu)
u_{11}^{(0)} \left[ \bar u_{42,e}^{(m)} \gamma_\beta g^Z(e) u_3 +
\bar u_4 \gamma_\beta g^Z(e) u_{23,e}^{(m)} \right] \\ \nonumber
%
&  & \mbox{} + \sum_{V=\gamma,Z} P_V^{\alpha \beta}(p_1-p_2-k_1) \,
\bar u_{21}^{(m)} \gamma_\alpha g^V(e) u_1 \left[
\bar u_4 \gamma_\beta g^V(e) u_{23,e}^{(m)} +
\bar u_{42,e}^{(m)} \gamma_\beta g^V(e) u_3 \right] \\ \nonumber
%
&  & \mbox{} + P_W^{\alpha \beta}(p_1-p_2-k_2) \,
\bar u_4 \gamma_\alpha g^W(e) u_{13}^{(0)} \left[
\bar u_2 \gamma_\beta g^W(e) u_{21,e}^{(m)} +
\bar u_{22,\nu}^{(0)} \gamma_\beta g^W(e) u_1 \right] \;,\\
%
%
\nonumber
{\cal M}^{(l-o)} & = & \sum_{V=\gamma,Z} -g_{VWW} P_V^{\alpha
\beta}(p_1-p_2-k_1)
\Gamma_\alpha(p_1-p_2,-k_1;J_1,\epsilon_1) \\ \nonumber
&& \mbox{\hspace{1in}} \times \left[
\bar u_{42,e}^{(m)} \gamma_\beta g^V(e) u_3 +
\bar u_4 \gamma_\beta g^V(e) u_{23,e}^{(m)} \right] \\ \nonumber
%
& & \mbox{}  -g_{ZWW} P_W^{\alpha \beta}(p_1-p_2-k_2)
\Gamma_\alpha(-k_2,p_1-p_2;\epsilon_2,J_1)  \,
\bar u_4 \gamma_\beta g^W(e) u_{13}^{(0)} \\ \nonumber
%
&  & \mbox{} + \sum_{V=\gamma,Z} -g_{VWW} P_W^{\alpha \beta}(p_1-p_2-k_2)
\Gamma_\alpha(-k_1,p_3-p_4;\epsilon_1,J_2^V) \\ \nonumber
&& \mbox{\hspace{1in}}  \times \left[
\bar u_{22,\nu}^{(0)} \gamma_\beta g^W(e) u_1 +
\bar u_2 \gamma_\beta g^W(e) u_{21,e}^{(m)} \right] \;,\\
\nonumber
{\cal M}^{(p-r)} & = & \sum_{V=\gamma,Z} \left[ \bar u_{21}^{(m)}
\overlay{/}{J}_2^V g^V(e)
 u_{21,e}^{(m)} +
\bar u_{212,e}^{(m,m)} \overlay{/}{J}_2^V g^V(e) u_1 +
\bar u_{221,\nu}^{(0,m)} \overlay{/}{J}_2^V g^V(e) u_1 \right] \\ \nonumber
%
&  & \mbox{} + \bar u_2 \overlay{/}{J}_2^Z g^Z(\nu) u_{211,\nu}^{(0,0)} +
\bar u_{22,\nu}^{(0)} \overlay{/}{J}_2^Z g^Z(\nu) u_{11}^{(0)} +
\bar u_2 \overlay{/}{J}_2^Z g^Z(\nu) u_{121,e}^{(0,m)} \;,\\
\nonumber
{\cal M}^{(t,u)} & = &  \bar u_{42,e}^{(m)} \overlay{/}{J}_1 g^W(e)
 u_{13}^{(0)} +
\bar u_4 \overlay{/}{J}_1 g^W(e) u_{213,\nu}^{(0,0)} +
\bar u_2 \overlay{/}{J}_1 g^W(e) u_{123,e}^{(0,m)}\;,\\
\nonumber
{\cal M}^{(w)} & = &  - g_{ZWW} P_W^{\alpha \beta}(k_1+k_2)
\Gamma_\alpha(k_2,k_1;\epsilon_2,\epsilon_1) \; \bar u_4 \overlay{/}{J}_1
g^W(e)
\frac{ \overlay{/}{p}_3 - \overlay{/}{k}_1 - \overlay{/}{k}_2 }
     { (p_3-k_1-k_2)^2 }
\gamma_\beta g^W(e) u_3 \; ,\\
\nonumber
{\cal M}^{(x)} & = &  - g_{ZWW} P_W^{\alpha \beta}(k_1+k_2)
\Gamma_\alpha(k_2,k_1;\epsilon_2,\epsilon_1) \; \bar u_2 \overlay{/}{J}_2^Z
g^Z(\nu)
\frac{ \overlay{/}{p}_1 - \overlay{/}{k}_1 - \overlay{/}{k}_2 }
     { (p_1-k_1-k_2)^2 }
\gamma_\beta g^W(e) u_1 \;,\\
\nonumber
{\cal M}^{(y)} & = & \sum_{V=\gamma,Z} - g_{ZWW} P_W^{\alpha \beta}(k_1+k_2)
\Gamma_\alpha(k_2,k_1;\epsilon_2,\epsilon_1) \; \bar u_2 \gamma_\beta g^W(e)
\frac{ \overlay{/}{p}_2 + \overlay{/}{k}_1 + \overlay{/}{k}_2 + m}
     { (p_2+k_1+k_2)^2 - m^2}
\overlay{/}{J}_2^V g^V(e) u_1 \; .
\end{eqnarray}
The matrix elements for $e^+ e^- \rightarrow \bar \nu e^- W^+ Z$
can be obtained from those above by $CP$ conjugation.
%


\begin{table}
\caption{Cross sections in fb for the various channels contributing to the
$\nu\bar\nu VV$ Higgs-boson signals and backgrounds, demanding $|y(V)|<1$,
$p_T(VV)>45$~GeV, and vetoing central leptons $\bigl($with
$E_{\ell^\pm}>50$~GeV and $|\cos \theta_{\ell^\pm}|<\cos(0.15)\bigr)$.
Effects of beamstrahlung and
bremsstrahlung are not included here.  The row $\nu \bar \nu WW(0)$
describes continuum production, simulated by the choice $m_H=50$ GeV;
subtracting this from the $\nu\bar\nu WW$ row leaves the corresponding
Higgs-boson
signal. In the $\nu\bar\nu ZZ$ channel the Higgs-boson signal is about 1/3 of
the given
cross section. \label{various-ex}}
%
\begin{tabular}{lccccc}
&  $m_H=175$~GeV & $m_H=200$ & $m_H=250$ & $m_H=300$ & $m_H=350$ \\
\hline
$\nu \bar \nu WW$ & 42 & 29 & 16  & 9.8  &  6.9  \\
$\nu \bar \nu WW(0)$ & 5 & 5 &5 & 5  &  5  \\
$\nu \bar \nu ZZ$  & 1.3  & 13 & 8.8 & 5.4 & 2.7 \\
$e^+e^-WW$            & 0.15 & 0.15 & 0.14 & 0.13 & 0.12 \\
$e^+e^-ZZ$            & 0.016 & 0.014 & 0.01 & 0.009 & 0.009 \\
$e \nu WZ$            & 13  & 12  & 9.4 & 8.5 & 7.4  \\
$t\bar t$ ($m_t=150$ GeV) &  16 & 16 & 16 & 16 & 16
\end{tabular}
\end{table}
\begin{table}
\caption{Cross sections in fb for $\mu^+\mu^-VV$ channels, after cuts
$|y(V)|,\ |y(\mu^+\mu^-)|<1,\ \left|m(\mu^+\mu^-)-M_Z\right|<15$~GeV,
 $\overlay{/}{p}_T < 40$~GeV.
Effects of beamstrahlung and bremsstrahlung are not included here.
The row $\mu\mu WW(0)$ describes continuum production in this channel,
simulated by the choice $m_H=50$ GeV.
In the $\mu\mu ZZ$ channel, the continuum background cannot be
simulated in this way; it contributes about 2/3 of the integrated cross section
while the remaining 1/3 comes from the Higgs signal.
\label{mumu-ex}}
\begin{tabular}{lccccc}
&  $m_H=175$~GeV & $m_H=200$ & $m_H=250$ & $m_H=300$ & $m_H=350$ \\
\hline
$\mu \mu WW$ & 2.2 & 1.5 & 1.2  & 1.0  & 0.82  \\
$\mu \mu WW(0)$ & 0.64 & 0.64 & 0.64  & 0.64 & 0.64    \\
$\mu \mu ZZ$ & 0.11 & 1.2 & 0.86 & 0.56 & 0.28 \\
$t\bar t$ ($m_t=150$ GeV) & 0.26 & 0.26 &0.26 & 0.26 & 0.26 \\
\end{tabular}
\end{table}

\begin{table}
\caption{Cross sections in fb for the various channels contributing to the
$\nu\bar\nu VV$ Higgs signals and backgrounds, demanding $|y(V)|<1$,
$p_T(VV)>45$~GeV, and vetoing central leptons $\bigr($with
$E_{\ell^\pm}>50$~GeV and $|\cos \theta_{\ell^\pm}|<\cos(0.15)\bigr)$.
The effects of beamstrahlung, for Palmer~G, and bremsstrahlung are
included.
The row $\nu \bar \nu WW(0)$
describes continuum production in this channel, simulated by the choice
$m_H=50$ GeV. In the
$\nu\bar\nu ZZ$ channel, the signal is about 1/3 of the given cross section.
\label{various-in}}
%
%
%
\begin{tabular}{lccccc}
&  $m_H=175$~GeV & $m_H=200$ & $m_H=250$ & $m_H=300$ & $m_H=350$  \\
\hline
$\nu \bar \nu WW$ & 31 & 18 &  8.9 & 5.0 & 3.5  \\
$\nu \bar \nu WW(0)$ & 2.9 & 2.9 & 2.9 & 2.9  &  2.9  \\
$\nu \bar \nu ZZ$  & 0.62  & 6.9 &  4.0 & 1.8 & 0.69 \\
$e^+e^-WW$            & 0.15 & 0.16 & 0.15 & 0.15 & 0.15 \\
$e^+e^-ZZ$            & 0.08 & 0.05 & 0.016 & 0.011 & 0.008 \\
$e \nu WZ$            & 13  & 9.2  &  6.6 & 4.8  & 4.1   \\
$t\bar t$ ($m_t=150$ GeV) & 31 & 31 & 31 & 31 & 31 \\
\end{tabular}
\end{table}

\begin{table}
\caption{Cross sections in fb for $\mu^+\mu^-VV$ channels, after cuts
$|y(V)|,\ |y(\mu^+\mu^-)|<1,\ \left|m(\mu^+\mu^-)-M_Z\right|<15$~GeV,
$\overlay{/}{p}_T < 40$~GeV.
The effects of beamstrahlung, for Palmer~G, and
bremsstrahlung are included.
The row $\mu\mu WW(0)$ describes continuum production in this channel,
simulated by the choice $m_H=50$ GeV. In the $\mu\mu ZZ$ channel, the signal
is about 1/3 of the given cross section.
\label{mumu-in}}
%
%
%
\begin{tabular}{lccccc}
&  $m_H=175$~GeV & $m_H=200$ & $m_H=250$ & $m_H=300$ & $m_H=350$ \\
\hline
$\mu \mu WW$ & 0.87 & 0.71 & 0.54  & 0.41  & 0.31  \\
$\mu \mu WW(0)$ & 0.26 & 0.26 & 0.26  & 0.26 & 0.26  \\
$\mu \mu ZZ$ & 0.19 & 0.57 & 0.42 & 0.24  & 0.092  \\
$t\bar t$ ($m_t=150$ GeV) & 0.23 & 0.23 &0.23 & 0.23 & 0.23 \\
\end{tabular}
\end{table}

\begin{table}
\caption{Cross sections in fb for the various channels contributing to the
$\nu\bar\nu VV$ Higgs signals and backgrounds, demanding $|y(V)|<1$,
$p_T(VV)>45$~GeV, and vetoing central leptons $\bigl($with
$E_{\ell^\pm}>50$~GeV and $|\cos \theta_{\ell^\pm}|<\cos(0.15)\bigr)$.
We include the effects of bremsstrahlung plus beamstrahlung for the
DESY/Darmstadt narrow-band design; results for the TESLA design are almost
indistinguishable from these.
The row $\nu \bar \nu WW(0)$
describes continuum production in this channel, simulated by the choice
$m_H=50$ GeV. In the
$\nu\bar\nu ZZ$ channel, the signal is about 1/3 of the given cross section.
\label{tablefive}}
\begin{tabular}{lccccc}
&  $m_H=175$~GeV & $m_H=200$ & $m_H=250$ & $m_H=300$ & $m_H=350$  \\
\hline
$\nu \bar \nu WW$     & 42  &  25 &  14  & 8.9  & 5.9  \\
$\nu \bar \nu WW(0)$  & 4.5 & 4.5   & 4.5    & 4.5  & 4.5  \\
$\nu \bar \nu ZZ$     & 0.83  & 9.5 &  6.1 & 3.5 & 1.6 \\
$e^+e^-WW$            & 0.15 & 0.16 & 0.16 & 0.16 & 0.15 \\
$e^+e^-ZZ$            & 0.02 & 0.02 & 0.015 & 0.011 & 0.011 \\
$e \nu WZ$            & 14  & 10  &  8.5 & 7.4  & 6.7   \\
$t\bar t$ ($m_t=150$ GeV) & 19  & 19 & 19 & 19 & 19  \\
\end{tabular}
\end{table}

\begin{table}
\caption{Cross sections in fb for $\mu^+\mu^-VV$ channels, after cuts
$|y(V)|,\ |y(\mu^+\mu^-)|<1,\ \left|m(\mu^+\mu^-)-M_Z\right|<15$~GeV,
$\overlay{/}{p}_T <40$~GeV.
We include the effects of bremsstrahlung plus beamstrahlung for the
DESY/Darmstadt narrow-band design; results for the TESLA design are almost
indistinguishable from these.
The row $\mu\mu WW(0)$ describes continuum production in this channel,
simulated by the choice $m_H=50$ GeV. In the $\mu\mu ZZ$ channel, the signal
is about 1/3 of the given cross section.
\label{tablesix}}
\begin{tabular}{lccccc}
&  $m_H=175$~GeV & $m_H=200$ & $m_H=250$ & $m_H=300$ & $m_H=350$ \\
\hline
$\mu \mu WW$ & 1.8 & 1.4 & 1.1  & 0.87  & 0.7  \\
$\mu \mu WW(0)$ & 0.55 & 0.55 & 0.55  & 0.55 & 0.55  \\
$\mu \mu ZZ$ & 0.13 & 1.0 & 0.8 & 0.51  & 0.24  \\
$t\bar t$ ($m_t=150$ GeV) & 0.26 & 0.26 &0.26 & 0.26 & 0.26 \\
\end{tabular}
\end{table}

\begin{table}
\caption{Beamstrahlung and bremsstrahlung-corrected total cross sections in
fb of $e^+e^- \rightarrow\nu\bar\nu VV$, $\mu^+\mu^- VV$, where $VV=WW$, $ZZ$,
for various NLC designs, assuming $m_H=200$ and 300~GeV.
Also considered are the major background channels,
$e^+e^- \rightarrow e \nu WZ$, $t\bar t$.
For comparison, the results without QED corrections are given, too.
The same cuts are applied as in Tables III and IV.
\label{tableseven}}
\begin{tabular}{lcccc}
                       & No QED    & Palmer G & DESY/Darmstadt  & TESLA  \\
\hline
$\sigma\left(e^+e^-\to\nu\bar\nu VV\right)$ \\
\qquad $m_H=200$ GeV      &  42  & 25  & 35  & 36   \\
\qquad $m_H=300$ GeV      &  15  & 6.8  & 12  & 13   \\
\hline
$\sigma\left(e^+e^-\to e \nu WZ \rightarrow {\rm ``}\nu\bar\nu VV{\rm "}
\right)
$ \\
\qquad $m_H=200$ GeV      &  12  & 9.2  & 10  & 10   \\
\qquad $m_H=300$ GeV      &  8.5  & 4.8  & 7.4  & 7.4   \\
\hline
$\sigma\left(e^+e^- \to t\bar t \to {\rm ``}\nu\bar\nu VV{\rm "} \right)$ \\
\qquad $m_t=150$ GeV      &  16     &   31     &  19 & 19       \\
\hline
\hline
$\sigma\left(e^+e^- \to \mu^+ \mu^- VV\right)$  \\
\qquad $m_H=200$ GeV      &  2.7  & 1.3  & 2.4  & 2.4   \\
\qquad $m_H=300$ GeV      &  1.6  & 0.65  & 1.4 & 1.4   \\
\hline
$\sigma\left(e^+e^- \to t\bar t \to {\rm ``}\mu^+\mu^- VV{\rm "} \right)$ \\
\qquad $m_t=150$ GeV      & 0.26    &  0.23     &  0.26       & 0.26  \\
\end{tabular}
\end{table}


\figure{\label{ee-prod}
Feynman diagrams for Higgs-boson production in $e^+e^-$ collisions via (a)
$ZH$ associated production (with subsequent $Z$-boson decay)
and (b) $WW$ and $ZZ$ fusion.}

\figure{\label{s-depend}
Cross sections for Higgs-boson production in $e^+e^-$ collisions
versus CM energy $\sqrt s$, assuming (a) $m_H=100$~GeV and (b) $m_H=150$~GeV.
The solid curves represent $ZH$ associated production, allowing for $Z$-boson
virtuality below the nominal $ZH$ threshold.
The dashed (dot-dashed) curves represent $WW$ ($ZZ$) fusion.}

\figure{\label{branch}
Branching fractions versus $m_H$ for the decays of a Higgs boson into
(a) $WW$, $ZZ$, $Z\gamma$, and $\gamma\gamma$;
(b) $b \bar b$, $c \bar c$,  $\tau^+\tau^-$, and $gg$.
For the loop-induced decays $m_t=130$~GeV has been assumed.}

\figure{\label{e->nuW(Z)}
Generic lowest-order Feynman diagrams for the processes
$e^+e^- \rightarrow \ell_1 \ell_2 V_1V_2$.}

\figure{\label{distrib}
Differential cross sections $d\sigma/dm_{VV}$ for the production of two
leptons plus two weak bosons versus the $VV$ invariant mass $m_{VV}$,
where $V=W$ or $Z$, for the case $m_H=200$~GeV.
No cuts are imposed apart from the dimuon mass requirement of
Eq.~(\ref{mumu-cut}) in $\mu\mu VV$ channels and the fake-$V$ mass
requirement of Eq.~(\ref{eq8}) in $t\bar t$ background cases.
All contributions from both scattering and annihilation processes are included:
(a) $\nu\bar\nu VV$ signal and background channels, (b) $\mu^+\mu^- VV$
channels.}

\figure {\label{mh-nocut}
Total cross sections versus $m_H$: (a) $\nu\bar\nu VV$ signal and background
channels without cuts, (b) $\mu^+\mu^- VV$ channels with
$\left|m(\mu\mu)-M_Z\right|<15$~GeV.  The cross section
for $e^+e^- \rightarrow e^+e^-W^+W^-$ is relatively large so one fifth of it is
shown.  The fake-$V$ mass requirement of Eq.~(\ref{eq8}) is imposed in the
$t\bar t$ background cases.  The cross sections corresponding to four jets
in the final state may be obtained by multiplication with
$[B(V\rightarrow jj)]^2=0.5$.}

\figure{\label{PTmiss}
Differential cross sections without cuts for
(a) $e^+e^- \rightarrow \nu \bar \nu W^+W^-$ and
(b) $e^+e^- \rightarrow \nu \bar \nu H$  versus the missing
transverse momentum $\overlay{/}{p}_T$, assuming $m_H=200$~GeV. Scattering
and annihilation contributions are shown separately in (a);
$WW$-fusion and $ZH$-production are shown separately in (b).
The contributions due to annihilation channels
are summed over three neutrino flavors.}

\figure{\label{dist-cut}
Contributions to the $\nu\bar\nu VV$ signals and backgrounds, with acceptance
cuts but without initial-state radiative corrections, for the case $m_H=200$
GeV.
Differential cross sections are shown for $e^+e^-\rightarrow\nu\bar\nu
(e^+e^-,e\nu)VV$
versus the $VV$ invariant mass $m_{VV}$, where $V=W,Z$;  here $|y(V)|<1$,
$p_T(VV)>45$~GeV, and central $e^\pm$ vetoing by $E_{e^\pm}>50$~GeV and
$|\cos(\theta_{e^\pm})|<\cos(0.15)$ have been applied.
These curves receive contributions from both scattering and annihilation
channels. The background from $e^+e^- \to t\bar t$ production is also shown,
with the fake-$V$ mass constraint of Eq.~(\ref{eq8}) plus the corresponding
$p_T(VV)$ and $|y(V)|$ cuts and a veto on all central leptons
$e$, $\mu$, $\tau$.}

\figure {\label{mh-cut}
Total cross sections for the $\nu\bar\nu VV$ signals and backgrounds versus
$m_H$ from $e^+e^- \rightarrow \nu \bar \nu W^+W^-$, $\nu\bar\nu ZZ$,
$e \nu W^\pm Z$, and $t\bar t$, imposing the same cuts as in
Fig.~\ref{dist-cut}